\documentclass{elsart}
\usepackage{amssymb}
\usepackage{natbib}
\usepackage{graphicx}

\renewcommand{\theequation}{\thesection.\arabic{equation}}
\newcommand{\BibTeX}{ \textrm{B\kern-.05em\textsc{i\kern-.025em b}\kern-.08em
    T\kern-.1667em\lower.7ex\hbox{E}\kern-.125emX} }

\begin{document}

\begin{frontmatter}

% Title, authors and addresses

% use the thanksref command within \title, \author or \address for footnotes;
% use the corauthref command within \author for corresponding author footnotes;
% use the ead command for the email address,
% and the form \ead[url] for the home page:
% \title{Title\thanksref{label1}}
% \thanks[label1]{}
% \author{Name\corauthref{cor1}\thanksref{label2}}
% \ead{email address}
% \ead[url]{home page}
% \thanks[label2]{}
% \corauth[cor1]{}
% \address{Address\thanksref{label3}}
% \thanks[label3]{}

\title{Planck, Photon Statistics, and Bose-Einstein Condensation}

% use optional labels to link authors explicitly to addresses:
% \author[label1,label2]{}
% \address[label1]{}
% \address[label2]{}

\author[green]{Daniel M. Greenberger},
\author[scul]{Noam Erez},
\author[scul,svid]{Marlan O. Scully},
\author[scul,svid]{Anatoly A. Svidzinsky},
and
\author[scul]{M. Suhail  Zubairy}

\address[green]{City College of New York, New York, NY 10031}
\address[scul]{Institute~for~Quantum~Studies~and~Dept.~of~Physics,~Texas~A%
\&M~Univ.,~Texas~77843}
\address[svid]{Applied~Physics~and~Materials~Science~Group,~Eng.~Quad.,
Princeton~Univ.,~Princeton~08544}

%% This copyright statement isn't required at any stage by the Icarus
%% Editorial Office or Elsevier.  However, until you sign over the
%% copyright to Elsevier prior to publication (or negotiate with them
%% about copyright), you own the copyright to anything you create.
%% Just to keep things unambiguous, always include a copyright statement
%% or explicitly dedicate your work to the public domain.

\begin{abstract}

The interplay between optical and statistical physics is a rich and exciting field
of study.
Black body radiation was the first application of photon statistics,
although it was initially treated as a problem of the cavity oscillators
in equilibrium with the photon field.  However Planck surprisingly resisted
the idea that anything physical would be quantized for a long time after he
had solved the problem.  We trace this development.

Then, after the invention of the laser itself, it proved difficult
to develop a theory of laser action that could account for photon
statistics, i.e. fluctuations near threshold. This was accomplished
in 1965.  After Bose-Einstein condensation was successfully
achieved, the same problem arose in this case. The fluctuation problem
had not been treated adequately even for the ideal Bose gas. However this problem has
now been solved using the same techniques as in the theory of laser action.

\end{abstract}

\end{frontmatter}

\section{Introduction}

Optics was the original handle by which classical physicists learned to pry
their way into the mysteries of quantum physics. This was appropriate
because optics possesses a dual character, in one limit the purely classical
wave theory, and in the other the purely quantum mechanical particle limit.
In 1900, when the spectrum of black body radiation was being studied in
detail, only the classical side was known. This was used to connect it with
Thermodynamics, from which many of its properties could be derived. But the
Wien spectral law, which characterized the general, but not specific form of
the spectral law, was as far as thermodynamics could take one.

In order to get a specific law, Planck had to also draw on the probabilistic
considerations of Boltzmann, a real departure for Planck, and he
inadvertently drew into focus the particle aspect of the problem, without at
that time understanding just how radical his innovation was. But this added
statistics and fluctuations into the mix. A main point in this paper will be
to show the role that fluctuations played in Planck's and Einstein's
thinking in the early days of quantum theory, the important role it played
in the development of the quantum theory of the laser, and finally, how the
laser theory allows one to treat the fluctuations in a Bose-Einstein gas,
both above and below the critical temperature.

A second major theme in the paper will be to pursue the historical thread
running through Planck's work. In their desire to present a coherent story,
leading from classical physics to quantum physics, most textbooks leave out
or distort the history of the subject, which is consequently not well known.
But in this case, the courage of Planck in abandoning his lifelong distrust
of probability, coupled with his total reluctance to abandon the principles
of classical physics, led to a series of fascinating ironies that strongly
affected the history of the subject, and they deserve to be better known.

A further hidden element guiding the development of the early quantum
theory, the laser, and Bose-Einstein condensation, was the connection
between advancing technology and experimental technique. The effects of
technology are apparent in the laser and Bose-Einstein condensation,
although they are usually not appreciated as an input into early quantum
theory, but the accurate measurements of the black body spectra were made
possible by the invention of the bolometer by Langley, who became the first
director of the Smithsonian Institution in America. But also, the funding
for the improvements of the bolometer so that measurements could be extended
into the infrared, which became the most relevant measurements, leading to
the breakdown of Wien's specific radiation law, was provided by the power
company of Berlin, which city had recently been electrified. Black body
radiation is the least efficient means of illumination (one wants to be far
from equilibrium) and it set a standard against which to measure
efficiencies. So it turns out that the interest in funding such an abstruse
subject as black body radiation was actually driven by the technology of the
day.

The interconnections between all these threads forms an interesting subject
in itself, but here we shall only go so far as to follow a few of them. We
shall emphasize some of the interesting historical details that surround
Planck's work, which seem to be almost unknown to physicists. We have drawn
heavily on Planck's original papers, [Planck (1900 a,b)], reproduced with
comments in [Ter Haar (1967)], and Planck's book on heat radiation [Planck
(1913)]. We have also extensively used [Kuhn (1978)], [Hermann (1971)],
[Jammer (1966)], and [Heilbron (1996)]. Some other good references are
[Klein (1975)], [Mehra and Rechenberg (1982)], [Rosenfeld (1936)], [Varro
(2006)], and [Kangro (1976)]. An anthology that contains reprints of some
papers we refer to, with comments, is [Brush (2003)].

We go on to describe in some detail exactly what Planck did, and did not do,
and the importance of fluctuations in his work. Their true meaning and
importance was established by Einstein [Einstein (1909)]. We then describe
how fluctuations enter into the theory of the laser, and how this theory has
been used to treat fluctuations in Bose-Einstein Condensation.

\section{Planck's Black Body Radiation Law}

\setcounter{equation}{0}

\subsection{Some Ironical Historical Details Concerning Planck}

It is rather universally assumed that when Planck introduced the quantum in
1900, [Planck (1900a), Planck (1900b)] he quantized the energy levels of an
oscillator. But in fact, what he did was very ambiguous [Kuhn (1978)], and
we shall produce some strong evidence that at that time he was thinking more
along the lines of quantizing the size of cells in phase space. Furthermore,
he fought the idea of quantizing both radiation and the oscillator. In fact,
as late as 1913, when he published the second edition of his book on heat
radiation [Planck (1913)], he did not believe that the energy levels of
either the oscillator or the radiation were quantized, even though Einstein
had quantized both of them, the photon in 1905, [Einstein (1905)] and the
oscillator in 1907 [Einstein 1907]. We shall introduce a number of quotes
from Planck's original theoretical paper on quantum theory [Planck (1900b)]
which is usually taken as the birth of quantum theory, and from his heat
radiation book, to prove his aversion to quantizing anything physical. It
was not until Bohr had quantized the levels of the hydrogen atom [Bohr
(1913)], and the discussions that followed this, that Planck and most of his
colleagues accepted quantization as a fact of nature.

It is well known that Planck rejected the idea of photons until quite late,
but here is a quote from the introduction of his 1913 book that not only
proves that, but that also outlines his philosophy on the subject, which we
think not only explains his opinions, but also made it possible for him to
discover the law of black body radiation long before either he or anyone
else understood its consequences. He says, ``While many physicists, through
conservatism, reject the ideas developed by me, or at any rate maintain an
expectant attitude, a few authors have attacked them for the opposite
reason, namely, as being inadequate, and have felt compelled to supplement
them by assumptions of a still more radical nature, for example, by the
assumption that any radiant energy whatever, even though it travel freely in
a vacuum, consists of indivisible quanta, or cells. Since nothing probably
is a greater drawback to the successful development of a new hypothesis than
overstepping its boundaries, I have always stood for making as close a
connection between the hypothesis of quanta and the classical dynamics as
possible, and for not stepping outside of the boundaries of the latter until
the experimental facts leave no other course open. I have attempted to keep
to this standpoint in the revision of this treatise necessary for a new
edition." Obviously he is referring disapprovingly to Einstein's photons in
the quote. However we shall see that the last part of the quote is also very
relevant in deciding what he actually did.

What then led to the radically new form of the radiation law? It was the
breaking up of the energy cells into finite units, so that statistics could
be applied, which as he said he took directly from Boltzmann [Boltzmann
(1877)] (who ultimately let the cell size go to zero), and his introduction
of a new way of counting microstates. We point out that the introduction of
the quantum alone was not enough to produce his formula. This is because
when one uses Boltzmann statistics, if one changes the cell size, one merely
changes the thermodynamic probability by an exponential multiplicative
factor, which in turn leads to an additive constant in the entropy. Since in
classical entropy, an additive constant has no physical significance, this
is why the cell size doesn't matter in classical physics. Planck's argument,
which led to a counting scheme that looks very like Bose statistics,
introduced a cell size that is unique, and in fact is a fundamental constant
of nature, and this was caused by his taking the entities that occupied
these quantized units as indistinguishable. He was silent on this matter,
and it took a long time for people to realize it.

The first inkling of what was happening came from two papers in 1911, one by
Natanson [Natanson (1911)], and the other by Ehrenfest (1911)]. They both
singled out Einstein's derivation of the localized photon-like properties of
electromagnetic waves [Einstein (1905)] in the limit where Wien's radiation
formula worked (Einstein took Wien's formula as his starting point). They
pointed out that his argument would not work with Planck's formula instead
of Wien's and that Einstein's argument presumes Boltzmann statistics. They
then point out that Planck's argument assumes that the energy units are
indistinguishable, which they each find very puzzling. (Of course with
hindsight, we realize that Wien's formula holds in the particle-like domain
where Einstein was operating, while the Rayleigh-Jeans formula holds in the
wave-like regime. Einstein's later 1909 paper on fluctuations [Einstein
(1909)] sets out the particle-wave dichotomy for photons for the first time.)

Before we begin, we would like to point out that there were many historical
ironies in Planck's development. His thesis advisor (1879), Phillip von
Jolly, told him that the development of the first and second laws of
thermodynamics had completed the structure of theoretical physics and that a
bright young man should think twice about entering the field [quoted by
Heilbron (1996)]. (This in spite of the fact that Maxwell's equations had
been developed only ten years earlier. But it is known that Einstein in 1900
couldn't find a course on electrodynamics at Zurich, and had to teach
himself. New advances percolated at a slow rate in those days.)

Planck nonetheless thought that there was a lot left to do regarding
entropy, and spent most of his early career developing the consequences of
the second law, namely in chemical thermodynamics. This led to a call to
Berlin in 1889 for him to replace Kirchoff, who had retired (but Planck was
not initially appointed as a full professor). He soon complained in a letter
that ``nobody in Berlin is interested in entropy" [Heilbron (1996)]. But
when he started working on black body radiation, he immediately looked for
the connection between entropy and energy, while he said everyone else was
looking for the connection between frequency and Temperature.

Although he was initially under the influence of Ostwald and Mach (the great
disbelievers in atomic theory, since atoms were then considered
unobservable), Planck had slowly come to believe in atoms, as he thought it
was the only way to treat certain problems, such as heat conduction and
osmotic pressure, but he was sure that they were to be treated by mechanics.
He was bitterly against the probability arguments of Boltzmann, whom he
otherwise respected, because he thought the second law had to be exact. In
fact he set an assistant, Zermelo (later of axiomatic set theory fame), to
develop one of the two main arguments against Boltzmann, the ``ergodic"
argument [Zermelo (1896a,b), Boltzmann (1896a,b)], that a system in phase
space will ultimately return to a point arbitrarily close to where it is
now, even if far from equilibrium. (The other argument, due to Loschmidt,
was the ``time reversal" argument [Boltzmann (1877b)], that for every state
heading toward equilibrium, there is another time-reversed state heading
away from it).

He started to work on the black body radiation problem in 1896, and he
thought [Planck (1900c)] he had proven Wien's empirical radiation law (an
exponential form, which actually holds only at relatively low temperatures,
or high $\nu /T$). By 1900, experiments were being carried out at higher
temperatures and lower, infra-red frequencies, and the experimentalists,
Lummer and Pringsheim, and Planck's colleagues, Rubens and Kurlbaum, were
finding out that Wien's law did not work. The energy at a given frequency at
higher temperatures was becoming linear in the temperature (in accord with
the not-yet-stated Rayleigh-Jeans law).

Planck developed his radiation law in a somewhat ad-hoc manner, which law
worked very well, and he then set about to develop a theoretical
explanation. At that time, there were a number of proposed hypothetical laws
to deal with the discrepancy being discovered in Wien's law. Planck's worked
almost perfectly, and was quickly accepted by the experimentalists. But it
was clear that for any law to be taken seriously, it had to be theoretically
motivated. Planck had become convinced that one could never discover the
universal energy function by purely thermodynamic means, and he reluctantly
decided to switch to Boltzmann's methods. In a later famous letter that
Planck wrote about the success of that effort, [letter to R. W. Wood, quoted
in Hermann (1971)], he said that switching to probability arguments was ``an
act of desperation", but that breaking the cell into units of $h\nu $ was
``purely a formal assumption and I really did not give it much thought"
(since the dependence on $\nu $ is actually required by Wien's spectral law,
a direct consequence of the second law).

In order to apply the probability theory, Planck wrote $S=k_B\ln W$, to
connect the thermodynamic probability $W$ with the entropy $S$. In doing so,
he wrote this equation for the first time, as Boltzmann, on whose tombstone
the equation appears, never actually wrote it. Boltzmann always used the 
\textit{H}-theorem, or something equivalent. Another related irony is that
Boltzmann never wrote $k_B$ as a separate constant, but always used ($%
R/N_{0} $), the gas-constant per molecule. Planck's radiation law allowed
one to calculate $h$, $k_B$ and $N_{0}$ accurately for the first time, as
well as the electrical charge e, from the Faraday constant. As a result,
Planck thought it only simple justice that $k_B$ should be called Planck's
constant, or at least the Planck-Boltzmann constant, but it never happened.
The poor fellow was stuck with $h$!

In 1908, Arrhenius (who wielded tremendous influence) tried to convince the
Swedish Academy [quoted in Heilbron (1996)] to give Planck the Nobel Prize
because ``it has been made extremely plausible that the view that matter
consists of molecules and atoms is essentially correct... No doubt this is
the most important offspring of Planck's magnificent work." No mention of
the quantum of action. But Planck had to wait another 10 years, because
Lorentz [Lorentz (1908)] had come up with an argument that the
Rayleigh-Jeans law had to be the correct classical law, and that the reason
it failed at high temperature was that the system could not come to
equilibrium at high temperatures. He withdrew this opinion after the
experimentalists convinced him that if the Rayleigh-Jeans law were correct,
with its ultraviolet catastrophe, many substances would glow in the dark at
room temperature. But the Academy decided that the jury was still out on
Planck's work.

An even further irony is that Planck was ultimately convinced of the truth
of the new quantum theory by Nernst's Heat Theorem, that $C_{V} \rightarrow
0 $ as $T\rightarrow 0$. This implies that $W\rightarrow 1$, and $%
S\rightarrow 0$, with no additive constant, so that there is an absolute
minimum entropy reached at absolute zero. This is a purely probabilistic
argument, so far had Planck gone in changing his view. The ultimate irony is
that when Boltzmann committed suicide, from sickness and frustration with
all his critics, the University of Vienna offered Planck his chair. (Planck,
who loved Vienna and was a professional quality pianist, was tempted. But
his colleagues at Berlin managed to make it worthwhile for him to stay.)

Planck [Planck (1913)] praises Einstein's derivation of $C_{V}$, but he
never mentions his quantization of the energy levels of the oscillator to $%
E_{n}=nh\nu $, or uses it. He merely says it is beyond the scope of his
book, but it strikes us as rather strange that he chose not to further
comment on it, as it seems to disagree with Planck's interpretation.

\subsection{Thermodynamic Background Leading to the Radiation Law}

The concept of a black body was introduced by Kirchoff in 1860 [Kirchoff
(1860)]. In what follows, including Planck's Law, we are going to give a
rather self-contained argument that will not always be historically
complete, although we will indicate certain occasions where historic remarks
are relevant, because they determine motivations, and provide a context for
what people did. We note that the history is often fairly complicated,
controversies arose and sometimes took many years to get resolved. Sometimes
it is even true that no one knew precisely what had been accomplished until
much later. (As an example, we note that the ``ultraviolet catastrophe" was
not even named as such [Ehrenfest 1911] until 11 years after Planck had
solved the problem!) We are not professional historians, but at least in the
case of Planck, there are many ``smoking guns" within his work to justify
what we say.

Kirchoff knew from looking at spectral lines from the sun that there was
heat energy in empty space, and postulated equilibrium radiation. But the
knowledge of what it consisted of was primitive. Maxwell's equations had not
yet been postulated, and the identity of heat rays and light rays had not
yet been established. Nor had the existence of atoms in the walls of a
cavity, nor that an oscillator radiates and absorbs electromagnetic energy,
or that such energy carries momentum. Thus it is rather amazing that
Kirchoff should have established on the basis of relatively simple arguments
that within a cavity at equilibrium, this radiation should be independent of
the substance of the walls of the cavity, and that at a fixed temperature a
good emitter of radiation should be a good absorber. A perfect absorber
should then radiate an energy equivalent to everything that falls upon it
within the cavity at equilibrium, independently at each frequency. The
radiation emitted by such a perfect absorber he called black radiation, and
there should then be a universal function $u(\nu ,T)$ that describes the
radiation density in equilibrium with the walls, that on average gets both
absorbed and reemitted, at any particular frequency and temperature.

Because of the unknown nature of what happened within the cavity, Kirchoff
was attacked for each of the assumptions he made leading to this conclusion,
and the existence of this universal function was dismissed by many.
Meanwhile others tried to change the assumptions and re-derive the results.
Even after the turn of the century this argument went on (well after
Planck's work). Although Planck does not explicitly mention these
controversies in the 1913 edition of his book on heat radiation, he was
nonetheless clearly affected by them, as he takes over 20 pages to discuss
and justify Kirchoff's law.

However, after Maxwell, Boltzmann tried to find a thermodynamic ``equation
of state" for the radiation in 1884 (similar to $PV=NRT$ for particles)
[Boltzmann (1884)], and after Hertz had produced electromagnetic waves in
1888, Wien tried in 1893 [Wien (1893)] to find the spectral function of
Kirchoff. He succeeded to the extent of reducing the problem to a single
function of $\nu /T$, which is as far as one can go thermodynamically, and
for which he ultimately won the Nobel Prize.

Since the radiation hitting an area A of the wall of a cavity carries both
momentum density and energy density, Boltzmann was able to treat it
similarly to a particle flux hitting the wall, and showed that%
\begin{equation}
P(T)=\frac{1}{3}u(T)  \label{g2.1}
\end{equation}

where $P_{\nu }$ and $u_{\nu }$ refer to the pressure and energy density
between frequencies $\nu $ and ($\nu $ + $d\nu $). The difference between
this formula and the non-relativistic one is the factor 1/3, rather than the
2/3 for particles, which comes from the non-relativistic form for the energy
($E=mv^{2}/2=pv/2$) rather than the extreme relativistic form for light ($%
E=pc$). Boltzmann then used this in connection with the second law of
thermodynamics%
\begin{equation}
dU=TdS-PdV,\quad \left( \frac{\partial U}{\partial V}\right) _{T}=T\left( 
\frac{\partial S}{\partial V}\right) _{T}-P=T\left( \frac{\partial P}{%
\partial T}\right) _{V}-P,  \label{g2.2}
\end{equation}%
together with $U=Vu$, to get%
\[
u=T\left( \frac{1}{3}\frac{du}{dT}\right) -\frac{1}{3}u, 
\]%
\[
4u=T\frac{du}{dT}, 
\]

\begin{equation}
\ln u=4\ln T+const,\quad u=\sigma T^{4},  \label{g2.3}
\end{equation}%
the Stefan-Boltzmann law. Before Boltzmann derived it theoretically, Stefan
had correctly guessed its form by examining some data that was not only
inadequate, but that we now know was inaccurate as well.

For the entropy, we again use the equation (\ref{g2.2}). If we define the
entropy density $s$ as $S=Vs$, then%
\[
u=Ts-\frac{1}{3}u, 
\]

\begin{equation}
s=\frac{4}{3}\frac{u}{T},\quad S=Vs=\frac{4}{3}\sigma T^{3}V.  \label{g2.4}
\end{equation}%
Eq. (\ref{g2.4}) implies that during an adiabatic expansion of the cavity,
so that the entropy is constant, we will have $VT^{3}=const.$

\subsubsection{Wien's Spectral Law}

In 1888 Hertz showed the reality of Maxwell waves. In 1893 Wien applied the
laws of thermodynamics and electromagnetism to the problem of black body
radiation [Wien (1893)] and succeeded in reducing Kirchoff's universal
function to a function of one variable. That is as far as one can go in
classical physics. Wien tackled the problem of including the frequency in
the black body law by considering an adiabatic motion of a wall of the
cavity. This induced a Doppler shift on the radiation, while at the same
time the wall did work on the radiation. Born's Atomic Physics book [Born
(1929)] has a simplified treatment in an appendix. But we will consider a
much simpler technique based on adiabatic invariance, that was not available
to Wien, but was first introduced by Ehrenfest in 1913 [Ehrenfest (1913)].
Ehrenfest was looking for some quantity that would not change while the
external parameters of the system undergo a slow adiabatic change. He
reasoned that such a quantity would be a good candidate for quantization,
since it would not undergo a gradual change during the process, but could
only change abruptly. This became the theoretical underpinning for the
Bohr-Sommerfeld-Wilson quantization rule.

First we have to find the normal modes of the radiation. We assume the
cavity is a cube, of side L, since for all but the lowest normal modes the
shape does not matter. We also assume that the walls are fully reflecting
and use standing wave boundary conditions. Then the modes for a Fourier
expansion of the field satisfy%
\begin{equation}
k_{x}L=n_{x}\pi ,\quad k_{y}L=n_{y}\pi ,\quad k_{z}L=n_{z}\pi ,\quad
k^{2}=\omega ^{2}/c^{2}.  \label{g2.5}
\end{equation}%
The last of these equations comes from the wave equation for the fields.
Here the n's are positive integers. The number of modes in a region is given
by%
\begin{equation}
\Delta n_{x}\Delta n_{y}\Delta n_{z}=\Sigma _{n}=\frac{L^{3}}{\pi ^{3}}%
\Delta k_{x}\Delta k_{y}\Delta k_{z}\rightarrow \frac{L^{3}}{8\pi ^{3}}4\pi
k^{2}dk=\frac{L^{3}}{8\pi ^{3}}4\pi \frac{8\pi ^{3}}{c^{3}}\nu ^{2}d\nu .
\label{g2.6}
\end{equation}%
The $8$ in the denominator of the first line is due to the fact that the $n$%
's are positive, so only the first octant is important, but we are
integrating the $k$'s over all of $k$ space. Finally we must introduce
another factor of $2$ because there are two degrees of polarization for each
direction in $k$ space. So%
\begin{equation}
\Sigma _{n}=V\frac{8\pi }{c^{3}}\nu ^{2}d\nu .  \label{g2.7}
\end{equation}%
Rayleigh introduced the counting of modes of the field [Rayleigh (1900)] in
1900. He did it only qualitatively, following an earlier procedure he had
used for sound waves. He then said the total energy density should be%
\begin{equation}
u_{\nu }d\nu =dU/V=\Sigma _{n}\bar{\varepsilon}_{n}=\frac{8\pi }{c^{3}}\nu
^{2}k_{B}Td\nu .  \label{g2.8}
\end{equation}%
Here represents the average energy of a mode, which by the equipartition
theorem should be $k_{B}T$. In 1905 he added the numerical factors in the
above equation [Rayleigh (1905)], but made a minor mistake which was
corrected by Jeans [Jeans (1905)], who emphasized how important and
inescapable the above formula is. It has since been known as the
Rayleigh-Jeans Law. Later Lorentz also gave a very general derivation
[Lorentz (1908)], and for a while he and Jeans believed that the reason the
equation did not work experimentally was because it was difficult to
establish equilibrium at high frequencies, and the experiments were
therefore not correct. But the equation blows up at high frequencies and so
cannot be correct, a problem labeled by Ehrenfest as the ``ultraviolet
catastrophe" in 1911 [Ehrenfest (1911)].

To establish Wien's law, one need only note that in Eq. (\ref{g2.5}), if one
slowly changes $L$, then $k_{i}$ will slowly change, but $n_{i}$ cannot and
will stay fixed [Ter Haar (1967)]. This leads to%
\[
k_{i}L=const,\quad \nu L=const,\quad \nu ^{3}V=const, 
\]

\begin{equation}
T^{3}V=const,\quad (adiabatic \quad change)  \label{g2.9}
\end{equation}%
\[
\nu /T=const. 
\]%
The second line above is just Eq. (\ref{g2.4}), and so since the entropy of
each node, $s_{n}$, remains constant during an adiabatic change, one must
have%
\begin{equation}
S/V=\Sigma _{n}s_{n}=\frac{8\pi }{c^{3}}\int \nu ^{2}s_{n(\nu )}(\nu /T)d\nu
\equiv \int s_{\nu }d\nu ,  \label{g2.10}
\end{equation}%
and therefore%
\begin{equation}
u_{\nu }=\frac{3}{4}Ts_{\nu }=\frac{3}{4}\nu ^{3}(T/\nu )\frac{8\pi }{c^{3}}%
s_{n}(\nu /T)=\nu ^{3}f(\nu /T).  \label{g2.11}
\end{equation}%
The Rayleigh-Jeans law obviously takes this form, and so does an empirical
radiation law proposed by Wien, [Wien (1996)]%
\begin{equation}
u_{\nu }=a\nu ^{3}e^{-b\nu /T}.  \label{g2.12}
\end{equation}%
(We call this Wien's radiation law, to differentiate it from Eq. (\ref{g2.11}%
), Wien's spectral law, firmly embedded in the laws of thermodynamics. Eq. (%
\ref{g2.11}) is sometimes called Wien's displacement law, but we reserve
this for the statement concerning the frequency where the energy
distribution $u_{\nu }$ is a maximum, $\nu _{\max }$/T = const, a
consequence of Eq. (\ref{g2.11}).) Prior to 1900, all measurements were
taken in the relatively high frequency domain, and Wien's empirical law held
pretty well. In fact, Planck had convinced himself that it must be the
universal law. But the situation started changing after improvements were
made to the experimental equipment. Then Rubens reported to Planck that at
higher temperature for a given frequency the results were becoming linear in
T, and Planck realized he had to rethink his ideas.

\subsection{Planck's Introduction of the Quantum of Action}

In his first theoretical paper in 1900, Planck [Planck (1900b)] makes two
very confusing statements about the quantization of energy. He gives two
successive sentences that are totally contradictory. After telling us that
he will use Boltzmann's method, he says, ``If $E$ [the energy of the $N$
resonators of energy $\nu $] is considered to be a continuously divisible
quantity, this distribution is possible in infinitely many ways. We
consider, however - this is the most essential point of the whole
calculation - $E$ to be composed of a very definite number of equal parts
and use thereto the constant of nature $h=6.55\times 10^{-27}$erg$\cdot $%
sec. This constant multiplied by the common frequency $\nu $ of the
resonators gives us the energy element $\varepsilon $ in erg, and dividing $%
E $ by $\varepsilon $ we get the number $P$ of energy elements which must be
divided over the $N$ resonators." This statement is often quoted in history
of quantum theory books and articles, and it certainly looks like Planck is
talking about quantized energy levels.

However the very next sentence reads, ``If the ratio is not an integer, we
take for $P$ an integer in the neighborhood." Now if he really meant for the
energy units to be quantized, $P$ would naturally be an integer. Instead, we
believe that he meant that the resonators could have any energy between $%
n\varepsilon $ and $(n+1)\varepsilon $, and one just lumped them all
together as $n\varepsilon $. In other words, he was quantizing in phase
space, as Boltzmann had done, because as he said, one could not count states
otherwise. He went on to count and characterize the energy elements, $%
\varepsilon $, but he never said that an oscillator's total energy must be $%
n\varepsilon $, as Einstein later did. This is because, as we shall show
below, he never believed it to be so.

In 1906-7, Planck gave a series of lectures in Berlin, which were published
as a rather comprehensive book on ``heat radiation". He put out a second
edition in 1913, [Planck (1913)] So the statements in the book should be
indicative of how Planck thought about the subject as late as 1913.

There is no doubt that he introduced a quantum of action. He says as much in
opening the preface to the second edition, ``Recent advances in physical
research have, on the whole been favorable to the special theory outlined in
this book, in particular to the hypothesis of an elementary quantity of
action." But exactly what was quantized? He says on p. 125, ``By the
preceding developments the calculation of the entropy of a system of N
molecules in a given thermodynamic state is, in general, reduced to the
single problem of finding the magnitude $G$ of the region elements in the
state space. That such a definite finite quantity really exists is a
characteristic feature of the theory we are developing, as contrasted with
that due to Boltzmann, and forms the content of the so-called hypothesis of
quanta."

It would seem fairly certain from this statement that his interest was in
quantizing phase space. Shortly thereafter, in Part III, chapter III, p.
135, he introduces a model of the linear harmonic oscillator, specifically
in phase space. He talks about the energy as an ellipse, and makes the
transition from the coordinates $p$ and $x$ to $E$ and $\varphi $. He
introduces the unit of action and takes the ellipses to have the average
energy $(n+1/2)\varepsilon $. He then makes an argument defending the
appearance of what we now call ``zero-point energy" (although his
interpretation of it is totally different, having nothing to do with the
uncertainty principle). It is hard to see why he would do that unless he
thought the actual energies were distributed throughout the ellipse.

In part IV, chapter III, he shows in more detail his ideas about the
emission of radiation. To modern eyes, this new theory of Planck's looks
very strange, as it makes absorption and emission totally different
processes. But it was used by a number of people for a while, and it could
explain the photoelectric effect, and a few other things, but it was
forgotten relatively soon after Bohr quantized the Hydrogen atom later that
year. (Bohr's theory itself took some time to become accepted.) But it shows
how Planck's thinking was totally in flux, and how even then he was
unwilling to believe in the quantization of energy levels. On p. 161, he
says, ``Whereas the absorption of radiation by an oscillator takes place in
a perfectly continuous way, so that the energy of the oscillator increases
continuously and at a constant rate, for the emission we have, in accordance
with sec. 147, the following law: The oscillator emits in irregular
intervals, subject to the laws of chance; it emits, however only at a moment
when its energy of vibration is just equal to an integral multiple $n$ of
the elementary quantum $\varepsilon $ $=h\nu $, and then it always emits its
whole energy of vibration $n\varepsilon $."

He then describes how the oscillator absorbs energy at a constant rate, so
that its energy increases linearly in time, and as it passes a given energy $%
n\varepsilon $, it may or may not radiate. If not, it continues on toward $%
(n+1)\varepsilon $. So the oscillator energy is not quantized, but it emits
in quantized units, of multiples of the quanta. On the basis of this model,
he then goes on to calculate, p. 166, ``Hence in the state of stationary
equilibrium the number of oscillators whose energy lies between $nh\nu $ and 
$(n+1)h\nu $ is..." and proceeds to give a complicated formula. But it is
clear that the energy levels of the oscillator are not quantized, nor is the
absorption of radiation. Only the emission of radiation is. Presumably after
emission, the radiation got thermalized. So by this time in his thinking,
something was quantized, but it did not stay quantized. He even draws a
diagram giving the saw-toothed form described above for the energy of a
single oscillator as a function of time.

We would like to say something about Planck's intellectual attitude, which
was summarized in the quote we gave at the beginning. He was an insider, an
intellectual leader of the German community, and a man of total integrity.
He had not the slightest desire to overthrow, or to see the overthrow, of
the hard-won victories of classical science. And yet in times of crisis, he
had the moral courage both to introduce a notion that he knew was radical,
and whose implications no one could comprehend at the time, and also to
suddenly abandon a strong belief that had sustained him throughout his
career until then, namely that statistical considerations could not play a
fundamental role in the understanding of physics at a profound level. The
quote shows clearly that he would willingly go as far as he thought he had
to go, but absolutely no further, and he lived up to this conviction.

For this reason, we believe that nobody but Planck could have made the
advance that he made, when he did. His first paper was a purely
phenomenological gimmick, which he made by performing his analysis in terms
of entropy. As he said, he had devoted his life to examining entropy, which
few people at the time took seriously. In his second paper he realized that
the gimmick of the first paper had to correspond to a fundamental finite
unit of action. But what that meant, nobody was prepared to say at that
time. His own explanations were fuzzy, arbitrary, and had many loopholes. We
think his revelation took the subject as far as it could have gone without a
deeper analysis, which after all would consume many years of work by many
people. In the total state of ignorance at that time, we think he did
exactly what he was mentally inclined to do. He took the subject as far as
it could go at that time, and no further. He introduced the quantum of
action, and it worked, but its significance was very obscure. However it is
important to realize that quantized energy levels, for both radiation and
matter, are features of nature. Quantized cells in phase space are artifacts
of theory. It is interesting that he was willing to accept the latter, which
he could hope to fix, but was not willing for a long time to accept the
former, which would invalidate most of classical physics.

His conservatism led him for many years to try to find a close-to-classical
explanation for what he had done, and he was strongly inclined against the
radical advances of others, which is why it was left for Einstein to
quantize both the oscillator levels and the electromagnetic field. On the
other hand, in subjects where radical ideas could immediately lead to clear
conceptual advances, he was quick to approve, and he was one of the earliest
supporters of relativity theory, and in fact spent most of his research time
between 1905 and 1908 trying to advance the theory, and convince his peers
of its validity.

Planck's position in the German Physical Society made his voice the primary
one in deciding what should be published in \textit{Annalen der Physik}, the
leading physics journal of the time, and his openness to radical new ideas,
such as Einstein's, is almost without parallel (one wonders whether a paper
such as Einstein's special relativity paper would get published in \textit{%
Physical Review} today?) He even allowed Einstein to publish his photon
paper, with which he strongly disagreed.

\subsection{Planck's Derivation of the Black Body Radiation Law}

When Planck attacked the problem of black body radiation, he realized that
since the results were independent of the nature of the material in the
cavity, one could use a simple model for the cavity. So he chose to consider
a damped harmonic oscillator as a model for the material in the walls. His
results are arrived at simply in Born's book [Born(1949)]. For absorption of
radiation, if one has an oscillator of natural frequency $\omega _{0}$, and
weak damping, $\gamma $, which is being driven at frequency $\omega $, the
equation of motion will be the real part of%
\begin{equation}
m\ddot{x}+m\gamma \dot{x}+m\omega _{0}^{2}x=E_{0x}e^{i\omega t}.
\label{g2.13}
\end{equation}%
Then when one compensates for the 3-dimensionality of the problem, and
assumes that $E_{0}$ represents the equilibrium radiation present at
temperature $T$, one finds that the power absorbed is%
\begin{equation}
\frac{dE_{abs}}{dt}=\frac{\pi \kappa e^{2}}{3m}u(\nu _{0}),  \label{g2.14}
\end{equation}%
where $\kappa =1/4\pi \epsilon _{0}$ in mks units, while the power radiated
is given by%
\begin{equation}
\frac{dE_{rad}}{dt}=\frac{2\kappa e^{2}\bar{a^{2}}}{3mc^{3}}=\frac{2\kappa
e^{2}\omega _{0}^{2}}{3mc^{3}}\bar{\varepsilon},  \label{g2.15}
\end{equation}%
where the term with $a$ is the average of the acceleration-squared, and $%
\bar{\varepsilon}$ represents the average energy of the oscillator.
Combining Eqs. (\ref{g2.14}) and (\ref{g2.15}) gives%
\begin{equation}
u_{\nu }=\frac{8\pi \nu ^{2}}{c^{3}}\bar{\varepsilon}.  \label{g2.16}
\end{equation}

Planck had this result well before Rayleigh had published his node-counting
argument. All Planck had to do was insert the equipartition result $k_{B}T$
for $\bar{\varepsilon}$, and he would have had the Rayleigh-Jeans formula
considerably before Rayleigh.

But he never did, and there has been considerable debate as to why. Could he
have not known about equipartition, since at this time he was an avid
attacker of the entire statistical mechanics enterprise? This would seem
very unlikely, as he was interested in specific heats, and would have known
about the Dulong-Petit law controversy [Dulong, Petit (1819)] (some solids
did and some did not have $U=3RTV$). Or was he aware of it but already had
no confidence in it, as Wien's empirical law, eq. (2.12), seemed to be
holding up nicely. We are unlikely to ever know.

At any rate in 1900 Planck found out that Wien's law was not holding up, and
he had to make a report to the Berlin physical society. From his long
experience in thermodynamics, he later said that he immediately started
searching for the solution in the relation between entropy and energy, while
everyone else was worried about the relation between $\nu $ and $T$. Planck
had derived a formula for the approach to equilibrium by an oscillator in a
black body cavity that had a small excess energy $\Delta U$ over its
equilibrium value [Planck (1900d)]. Then if its energy changed by $dU$, the
change in entropy of the entire system (oscillator plus field) would be%
\begin{equation}
dS_{tot}=\frac{3}{5}\frac{d^{2}S}{dU^{2}}dU\Delta U.  \label{g2.17}
\end{equation}%
So the function $\frac{d^{2}S}{dU^{2}}$ is clearly connected to fluctuations
about equilibrium, although at the time Planck was not thinking
statistically. It was Einstein in 1909 [Einstein (1909)] who clearly brought
out the direct meaning of this function as a statistical measure of
fluctuations. He inverted the formula $S=k_{B}\ln W$ to the form $%
W=e^{S/k_{B}}$. Then one can connect the entropy of an arbitrary state to
its probability. If W is a maximum for $S=S_{0}=S(E_{0})$, the maximum
entropy and minimum energy state, then very close to equilibrium we can write%
\[
S=S_{0}-\alpha (E-E_{0})^{2},\quad -\alpha =\frac{1}{2}\left( \frac{\partial
^{2}S}{\partial E^{2}}\right) _{0}, 
\]

\begin{equation}
W=e^{S_{0}/k_{B}}e^{-\alpha (E-E_{0})^{2}/k_{B}}.  \label{g2.18}
\end{equation}%
There is no linear term since $S_{0}$ is a maximum. If we then ask for the
energy fluctuations about equilibrium, we get%
\begin{equation}
<\Delta E^{2}>=<(E-E_{0})^{2}>=\frac{\int (E-E_{0})^{2}W(E)dE}{\int W(E)dE}=%
\frac{k_{B}}{2\alpha },  \label{g2.19}
\end{equation}%
So $1/\alpha =-2/\left( \frac{\partial ^{2}S}{\partial E^{2}}\right) _{0}$
is a measure of the energy fluctuation.

Planck's first derivation of his radiation formula was a purely numerical
manipulation. Nonetheless, it is very interesting because it is profoundly
and directly connected to fluctuations, in a way that Planck could not have
foreseen. He knew that entropy was the key to the problem, and he thought
the answer was directly related to the quantity $1/\alpha $ of Eqs. (\ref%
{g2.18}) and (\ref{g2.19}), which governed the return to equilibrium, via
Eq. (\ref{g2.17}). Until a few days earlier, when Rubens had come to him, he
thought that Wien's empirical law, Eq. (\ref{g2.12}), was the correct
solution to the problem. Using his own Eq. (\ref{g2.16}), together with Eq. (%
\ref{g2.12}), he wrote%
\begin{equation}
\bar{\varepsilon}=\frac{c^{3}}{8\pi \nu ^{2}}u_{\nu }=\frac{ca\nu }{8\pi }%
e^{-b\nu /k_{B}T},  \label{g2.20}
\end{equation}%
One could use this to express $S$ directly in terms of $E$ by eliminating $T$%
, since at constant $V$, $1/T=\left( \frac{\partial S}{\partial E}\right)
_{V}=\frac{\partial s}{\partial \bar{\varepsilon}}$, where $s$ is the
entropy per oscillator. Therefore from Eq. (\ref{g2.20}),%
\[
\frac{1}{T}=-\frac{1}{b\nu }\ln \frac{8\pi \bar{\varepsilon}}{c^{3}a\nu }=%
\frac{\partial s}{\partial \bar{\varepsilon}} 
\]%
\begin{equation}
\frac{\partial ^{2}s}{\partial \bar{\varepsilon}^{2}}=-\frac{1}{b\nu \bar{%
\varepsilon}}.  \label{g2.21}
\end{equation}%
This is the expression Planck had previously thought exact, and even that he
could derive it with some plausible assumptions.

The new knowledge given to him by Rubens, that at low frequencies in the
newly accessible infra-red region, $u_{\nu }\approx \nu ^{2}T$, as had just
been predicted by Rayleigh, he wrote as (using $u_{\nu }=A\nu ^{2}T$)%
\[
\bar{\varepsilon}=\frac{c^{3}}{8\pi \nu ^{2}}u_{\nu }=\frac{c^{3}}{8\pi }%
AT=k_{B}T, 
\]%
\begin{equation}
\frac{1}{T}=\frac{\partial s}{\partial \bar{\varepsilon}}=\frac{k_{B}}{\bar{%
\varepsilon}},\quad \frac{\partial ^{2}s}{\partial \bar{\varepsilon}^{2}}=-%
\frac{k_{B}}{\bar{\varepsilon}^{2}}.  \label{g2.22}
\end{equation}%
(The last equation of the first line is just the equipartition theorem,
which was used by Rayleigh, although not by Planck, to give the value of $A$%
.) Planck says he spent the next few days looking for an extrapolation
between these two extremes, that gave plausible behavior, and finally came
up with%
\begin{equation}
\frac{\partial ^{2}s}{\partial \bar{\varepsilon}^{2}}=-\frac{k_{B}}{\bar{%
\varepsilon}(\Delta +\bar{\varepsilon})},  \label{g2.23}
\end{equation}%
where $\Delta $ is independent of the temperature. In the limit $\bar{%
\varepsilon}\ll \Delta ,\quad $%
\begin{equation}
\frac{\partial ^{2}s}{\partial \bar{\varepsilon}^{2}}\rightarrow -\frac{k_{B}%
}{\bar{\varepsilon}\Delta },\quad \Delta =bk_{B}\nu \equiv h\nu ,
\label{g2.24}
\end{equation}%
where $h$ is a new physical constant. The fact that $\Delta $ must depend
linearly on $\nu $ comes from Wien's spectral theorem, a thermodynamic
necessity. In the other limit, $\bar{\varepsilon}\gg \Delta $, we have Eq. (%
\ref{g2.22}), the Rayleigh-Jeans law. We can of course integrate Eq. (\ref%
{g2.23}), to get Planck's formula, which is still valid today,%
\[
\frac{\partial ^{2}s}{\partial \bar{\varepsilon}^{2}}=-\frac{k_{B}}{\bar{%
\varepsilon}(\Delta +\bar{\varepsilon})}=-\frac{k_{B}}{\Delta }\left( \frac{1%
}{\bar{\varepsilon}}-\frac{1}{\Delta +\bar{\varepsilon}}\right) , 
\]%
\[
\frac{\partial s}{\partial \bar{\varepsilon}}=\frac{1}{T}=-\frac{k_{B}}{%
\Delta }\ln \frac{\bar{\varepsilon}}{\Delta +\bar{\varepsilon}},\quad \frac{%
\Delta +\bar{\varepsilon}}{\bar{\varepsilon}}=e^{\Delta /k_{B}T}=e^{h\nu
/k_{B}T}, 
\]%
\begin{equation}
\bar{\varepsilon}=\frac{h\nu }{e^{h\nu /k_{B}T}-1}.  \label{g2.25}
\end{equation}

\subsubsection{Planck's Theoretical Derivation}

As we have said, Planck imagined that there were a series of oscillators in
the walls, in equilibrium with the radiation. Since each oscillator reaches
equilibrium with the same frequency of radiation as the oscillator itself,
and those of different frequencies all behave independently, we can consider
each frequency independently. This had previously given rise to much
controversy, the problem being how independently behaving oscillators could
ever come to equilibrium. especially if one considered the walls of the
cavity to be perfectly reflecting. The prevailing opinion was that this was
an abstraction, and if one thought of a small lump of coal (that absorbed
all frequencies) as also being inside the cavity, it would force all
frequencies to come to equilibrium together.

Next he considered that for each frequency, if there were $N$ oscillators,
the total energy was divided into $P$ discrete units of size $\varepsilon $
= $h\nu $. As we have said, it doesn't matter whether one considers this to
be because the energy is quantized, or because one considers all the energy $%
E$ between frequencies $\nu $ and $\nu $ $+d\nu $ to be lumped together and
considered as $E/\varepsilon =P$ discrete units. Planck in any case was
psychologically not disposed to seeing the energy as quantized, and as we
have emphasized, long resisted it.

If Planck were to continue following Boltzmann, he could further divide this
into $n_{k}$ oscillators with energy $k\varepsilon $ so that%
\begin{equation}
\sum n_{k}=N,\quad \sum k\varepsilon n_{k}=E,  \label{g2.26}
\end{equation}%
and then find the distribution of $n_{k}$'s which has maximum probability.
But Planck stated that one didn't even have to go this far. He merely said
that most of the time the system will be very close to equilibrium, and the
rest constitute rare events that will hardly contribute, so he just took the
total number of possible ways to distribute the P units of energy over the N
oscillators. How many such ways are there?

A simple way to see this (due to Ehrenfest) is just to draw two vertical
bars, and randomly distribute $P$ circles, and $N-1$ other bars between
them. For example, the arrangement

\[
|oo||o|ooo|||oo|.......|oo|o| 
\]

would represent $2$ units of energy in the first box (oscillator), none in
the second, $1$ in the third, $3$ in the fourth, etc., altogether taking up $%
N$ boxes. How many possible such arrangements are there? There are $N-1+P$
objects, which we can distribute in $(N-1+P)!$ ways, and since the order of
the circles and bars do not matter, the total becomes%
\begin{equation}
W=\frac{(N-1+P)!}{(N-1)!P!}  \label{g2.27}
\end{equation}%
This is the total number of ways of distributing the energy amongst the
oscillators, and the overwhelming majority of such arrangements lie close to
equilibrium.

Planck next assumed that $N\gg 1$, $P\gg 1$, and $\ln N!\approx N\ln N-N$.
Thus%
\[
S=k_{B}\ln W\approx k_{B}[\ln (N+P)!-\ln N!-\ln P!]= 
\]%
\[
k_{B}[(N+P)\ln (N+P)-(N+P)- 
\]%
\begin{equation}
(N\ln N-N)-(P\ln P-P)]+k_{B}N\left[ \left( 1+\frac{P}{N}\right) \ln \left( 1+%
\frac{P}{N}\right) -\frac{P}{N}\ln \frac{P}{N}\right] .  \label{g2.28}
\end{equation}%
So the entropy only depends on the average number of energy units per
oscillator, $P/N$. Then since%
\begin{equation}
\bar{\varepsilon}=E/N=P\varepsilon /N,\quad P/N=\bar{\varepsilon}%
/\varepsilon .  \label{g2.29}
\end{equation}%
(We have been using $\bar{\varepsilon}$ to represent the average energy per
oscillator, while $\varepsilon $ is just $h\nu $, the energy unit.) So
finally,%
\begin{equation}
s=S/N=k_{B}\left[ \left( 1+\frac{\bar{\varepsilon}}{\varepsilon }\right) \ln
\left( 1+\frac{\bar{\varepsilon}}{\varepsilon }\right) -\frac{\bar{%
\varepsilon}}{\varepsilon }\ln \left( \frac{\bar{\varepsilon}}{\varepsilon }%
\right) \right] .  \label{g2.30}
\end{equation}%
Then, as before%
\[
\frac{1}{T}=\frac{\partial S}{\partial E}=\frac{\partial s}{\partial \bar{%
\varepsilon}}=\frac{k_{B}}{\varepsilon }\ln \frac{\bar{\varepsilon}%
+\varepsilon }{\varepsilon }, 
\]%
\[
\bar{\varepsilon}=\frac{\varepsilon }{e^{\varepsilon /k_{B}T}-1}, 
\]%
\begin{equation}
u_{\nu }=\frac{8\pi \nu ^{2}}{c^{3}}\bar{\varepsilon}=\frac{8\pi \nu ^{2}}{%
c^{3}}\frac{\varepsilon }{e^{\varepsilon /k_{B}T}-1}.  \label{g2.31}
\end{equation}%
This is Planck's derivation of his formula. If we take an extra derivative
of the first line of Eq. (\ref{g2.31}), we get%
\begin{equation}
\frac{\partial ^{2}s}{\partial \bar{\varepsilon}^{2}}=-\frac{k_{B}}{\bar{%
\varepsilon}(\varepsilon +\bar{\varepsilon})}.  \label{g2.32}
\end{equation}

This reduces to Planck's previous numerical formula, where we see that $%
\varepsilon $ plays the role of his constant $\Delta $, which was necessary
to make the formula work. If $\varepsilon \rightarrow 0$, we lose the
behavior of Wien's empirical formula at high energies, which is the limit in
which Einstein introduced the particle behavior of photons. Rosenfeld, in
writing a history of early quantum theory [Rosenfeld (1936)] claimed that
Planck probably worked backward from Eq. (\ref{g2.31}) to get the entropy
Eq. (\ref{g2.30}), from which he could guess the right combinatorial law for
W, Eq. (\ref{g2.27}), which appears in Boltzmann's original article,
[Boltzmann (1877a)].

\subsection{Some Comments on the Planck Derivation}

There are a number of things to notice about Planck's derivation, some of
which we have noted earlier. First, what does it mean to keep $\varepsilon $
finite, since for the case of particles using classical statistics, cell
size doesn't matter? We pointed out earlier that it was noticed
independently by Natanson and Ehrenfest in 1911 that the Planck derivation
treats all the energy elements as equivalent, so that it is clearly
different from Boltzmann's statistics, and in fact makes them
indistinguishable. Ehrenfest also showed in 1906 [Ehrenfest (1906)] that the
Planck derivation puts an extra constraint on the system that he said could
be satisfied in several ways, but that the most natural was to strictly
quantize the energy levels of the oscillators. Einstein actually did this
[Einstein (1907)] in 1907, in his famous specific heat paper.

A number of people noticed that since $\varepsilon $ = $h\nu $, one of the
basic assumptions of the theory cannot work. Eqs. (\ref{g2.29}) and (\ref%
{g2.31}) assume that $N$ and $P>>1.$ But for high enough frequencies at a
given temperature, $\varepsilon $ becomes quite large, and most of the
oscillators will be in their ground state. This is why equipartition breaks
down, since classical physics scales the frequency so that all frequencies
are equally important, and they all have the same average energy, $k_{B}T$.
The Planck formula correctly identifies the parameter $\varepsilon /k_{B}T$
as the important dividing line, but the assumptions of the derivation also
break down at high frequency. The Einstein derivation of 1907 (where the
energy levels are quantized, and the probability that a state $%
E_{n}=n\varepsilon $ is occupied is $P_{n}=A\exp (-n\varepsilon /k_{B}T)$),
does not suffer from this defect.

Once he had shown that the energy levels of the oscillator are quantized,
Einstein also realized that Eq. (\ref{g2.16}), connecting $u_{\nu }$ with
the average energy of an oscillator, $\bar{\varepsilon}$, is inconsistent,
since it was derived using a classical oscillator that absorbs and emits
energy continuously. But he thought the equation must be true on the
average. So it is clear why Planck's derivation left the situation in a
state of great confusion for a long time.

\subsection{Einstein's Fluctuation Argument}

In 1909, Einstein looked at the fluctuations in the Planck formula [Einstein
(1909)] and noticed a simple, but very deep relation. It was in this paper
that Einstein introduced Eq. (\ref{g2.19}) for the fluctuations. We can see
the result already from Planck's early ad-hoc derivation of his result, Eq. (%
\ref{g2.23}). and if we insert Eq. (\ref{g2.23}) into Eq. (\ref{g2.19}), we
get%
\begin{equation}
<\Delta E^{2}>=\frac{k_{B}}{2\alpha }=-\frac{k_{B}}{\frac{\partial ^{2}s}{%
\partial \bar{\varepsilon}^{2}}}=\bar{\varepsilon}\Delta +\bar{\varepsilon}%
^{2},  \label{g2.33}
\end{equation}%
If one believes that the energy levels of the oscillator are quantized, so
that $E_{n}=n\Delta =nh\nu $, as Einstein did, and $\bar{\varepsilon}=\bar{n}%
\Delta $, where $\bar{n}$ represents the average level of the oscillator,
one can also put this into the form,%
\begin{equation}
\frac{<\Delta E^{2}>}{\Delta ^{2}}=<(\Delta n)^{2}>=\bar{n}+\bar{n}^{2}.
\label{g2.34}
\end{equation}

This also holds true for the field excitations if one considers the field
modes to be oscillators. He then pointed out that in his paper on photons in
1905 (they were not explicitly called ``photons" until 1926 [Lewis (1926)]),
he had used the Wien radiation formula when he discussed the radiation as
resembling individualized excitations, and had shown how it resembled the
independent particles of a perfect gas. He then identified the first term
with the fluctuations of a group of independent particles, while the second
term must correspond to the fluctuations in a cavity of classical waves.
(This result was expressed in terms of the energy density of a small finite
volume of the cavity, via Eq. (\ref{g2.16}), but the justification was
essentially a dimensional argument, which said that in the classical limit
where $\Delta $ doesn't contribute, and one has nothing else with the
dimensions of energy, one needs $<\Delta E^{2}>$: $\bar{\varepsilon}^{2}$.
An explicit later calculation by Lorentz [Lorentz (1912)] proved the
result.) But the Eqs. (\ref{g2.33}) and (\ref{g2.34}) are exact and hold
even when one is not in either of the two classical limits represented by
particles or waves. And so this paper is generally taken as the birth of the
wave-particle duality that has perplexed physicists up to the present time.

\subsection{Einstein's A and B Coefficients}

In 1917, Einstein [Einstein (1917)] published his famous $A$ and $B$
coefficients paper. The paper was in two parts, the first of which discussed
energy transformations and rates of absorption and emission for the various
processes that go on in an atom or molecule in equilibrium with the
radiation in a cavity. The second part discusses momentum transfer during
these processes. This paper was very seminal in that it taught us how to
think about radiation. It is not only the starting point for laser physics,
but it also pretty much made the existence of energy levels essential,
showing how they lead naturally to Planck's radiation law. Einstein assumed
that the molecule could occupy only a discrete set of allowed states \{$%
Z_{n} $\} which had energies \{$\varepsilon _{n}$\}, and whose relative
probability of occupation at temperature $T$ is%
\begin{equation}
W_{n}=p_{n}e^{-\varepsilon _{n}/k_{B}T},  \label{g2.35}
\end{equation}

where the $p_{n}$ represent statistical weights. He then assumes that a
molecule can decay spontaneously from a state $Z_{m}$ to $Z_{n}$, (such that 
$\varepsilon _{m}>\varepsilon _{n}$), and emit energy $\varepsilon
_{m}-\varepsilon _{n}.$ The probability per molecule for this to occur in
time $dt$ he takes as%
\begin{equation}
dW=A_{m}^{n}dt.  \label{g2.36}
\end{equation}%
As analogies, he quotes radioactive $\gamma $ decay and Hertzian oscillators.

He then assumes that there are induced (stimulated) emission and absorption
processes, which he calls a quantum theoretical hypothesis, that he assumes
take place with a probability%
\begin{equation}
dW=B_{n}^{m}u_{\nu }dt,  \label{g2.37}
\end{equation}%
for absorption from the lower level to the higher level, and%
\begin{equation}
dW=B_{m}^{n}u_{\nu }dt,  \label{g2.38}
\end{equation}%
for emission from the higher level to the lower level. These are for
transitions induced by the external field. Even without introducing the
quantized states, in the classical picture for absorption and emission used
by Planck, Eqs. (\ref{g2.14}) and (\ref{g2.15}), the rates were proportional
to the density of the surrounding radiation.

If we then equate emission and absorption at equilibrium (detailed balance),
we get%
\begin{equation}
p_{n}e^{-\varepsilon _{n}/k_{B}T}B_{n}^{m}u_{\nu }=p_{m}e^{-\varepsilon
_{m}/k_{B}T}(B_{m}^{n}u_{\nu }+A_{m}^{n}).  \label{g2.39}
\end{equation}

Then if we take the limit $T\rightarrow \infty $, for which also $u_{\nu
}(T)\rightarrow \infty $, then $p_{n}B_{n}^{m}=p_{m}B_{m}^{n}$, and%
\begin{equation}
u_{\nu }=\frac{A_{m}^{n}/B_{m}^{n}}{e^{(\varepsilon _{m}-\varepsilon
_{n})/k_{B}T}-1}.  \label{g2.40}
\end{equation}

This formula immediately leads to the Bohr rule $\varepsilon
_{m}-\varepsilon _{n}=h\nu $, and in the high temperature limit, where the
Rayleigh-Jeans law holds, we can evaluate $A/B$, which leads to the Planck
radiation law. Even after the development of non-relativistic quantum
mechanics, until the advent of field theory, Einstein's derivation was
needed to calculate the spontaneous emission of radiation.

Like in much of the rest of this story, there is an irony in Einstein's
introduction of his $A$ and $B$ coefficients. To Einstein himself, the most
important part of the paper was the second part. The derivation in this part
is more difficult, and is usually ignored today, but the point of the
calculation was the consideration of momentum conservation in the radiation
process, rather than merely energy conservation. By methods reminiscent of
his derivation of Brownian motion, he proved that to preserve thermal
equilibrium in a gas of atoms, or molecules, during the decay process one
must consider that in the individual decays, the atom recoils, acquiring the
appropriate momentum. In his words, ``If the molecule undergoes a loss of
energy of magnitude $h\nu $ without external influence, by emitting this
energy in the form of radiation (spontaneous emission), this process is also
a directed one. There is no emission in spherical waves. The molecule
suffers in the spontaneous elementary process a recoil of magnitude $h\nu /c$
in a direction which is in the present state of the theory determined only
by 'chance'."

The irony implicit in this derivation is brought out in his subsequent
statement, ``These properties of the elementary processes required by Eq.
(12) [an equilibrium equation of the momentum fluctuations] make it seem
practically unavoidable that one must construct an essentially quantum
theoretical theory of radiation. The weakness of the theory lies, on the one
hand, in the fact that it does not bring any nearer the connection with the
wave theory and, on the other hand, in the fact that it leaves moment and
direction of the elementary processes to 'chance'; all the same, I have
complete confidence in the reliability of the method used here."

This is the paper that introduced chance into the radiation process in an
essential way. After this, it was an inevitable and inescapable part of the
quantum landscape. He had to introduce it in order to make it clear that the
photons were emitted in individual quantum processes, and carried both
energy and momentum. This was very important to him, because even at this
late date, which was already 1917, the existence of the photon was not yet
generally accepted. But even as he introduced the element of chance in an
essential way, he lamented it.

There is a strong parallel here between Einstein and Planck, who both
introduced revolutionary thoughts brought about by necessity after a long
intellectual odyssey. Yet no sooner had Planck let the genie of the quantum
out of the bottle, than he devoted many years effort to unsuccessfully
trying to force it back in, without destroying the revolution it had brought
about. And Einstein had the same experience. Once he had let the genie of
chance out of the bottle, he unsuccessfully spent the rest of his life
trying to stuff it back in. Not that this diminishes by one iota the
accomplishments of these two great men, but it does point up the ironies
that life has in store for the best of us.

\section{Bose-Einstein Condensation}

\setcounter{equation}{0}

In 1924, Bose made the seminal observation that it is possible to derive
Planck's radiation law from purely corpuscular arguments without invoking at
all the wave properties of light resulting from Maxwell's field equations.
The main ingredient in Bose's argument was the indistinguishability of the
particles in question and a new way of counting them --- now universally
known as ``Bose-Einstein statistics'' --- which pays careful attention to
what is implied by their being indistinguishable. In the case of light
quanta, an additional feature is that their number is not conserved, because
light is easily emitted and absorbed. Massive particles (atoms, molecules,
\dots), by contrast, are conserved and therefore, as Einstein emphasized
[Einstein (1924), Einstein (1925)], their indistinguishability has further
consequences, of which the phenomenon of Bose-Einstein condensation (BEC) is
the most striking one.

Bose-Einstein condensation has long been a fascinating subject and has
attracted renewed interest in light of successful experimental
demonstrations of BEC in dilute $He^4$ [Crooker et al. (1983), Chan et al.
(1988), Crowell et al. (1995)] and ultracold atomic gases [Anderson et al.
(1995), Bradley et al. (1995), Davis et al. (1995), Fried et al. (1998),
Miesner et al. (1998)]. Furthermore the production of ``coherent atomic
beams'', the so called atom laser [Mewes et al. (1997), Andrews et al.
(1997), Anderson and Kasevich (1998), Bloch et al. (1999)], and its relation
to the conventional laser is intriguing; as is the relation between the BEC
phase transition and the quantum theory of the laser [Scully and Lamb
(1966), Scully and Zubairy (1997)].

The physics of BEC is subtle with many pitfalls and surprises. For example,
Uhlenbeck criticized Einstein's arguments concerning the implied singularity
in the equation of state at the critical temperature $T_c$. Einstein's
results require that the thermodynamic limit be taken, i.e., the number of
particles $N$ and the volume $V$ are taken to be infinite with the density $%
N/V$ being finite. This however leaves the question of how best to think
about and define $T_c$ for finite mesoscopic systems.

A canonical ensemble, in which $N$ particles inside a trap can interact and
exchange energy with a thermal reservoir at temperature $T$, provides a
natural approach to BEC. This canonical ensemble approach is a useful tool
in studying BEC properties in the current experiments on cold dilute gases
[Anderson et al. (1995), Bradley et al. (1995), Davis et al. (1995), Han et
al. (1998), Ernst et al. (1998), Hau et al. (1998), Esslinger et al. (1998),
Anderson and Kasevich (1999), Miesner et al. (1998), Fried et al. (1998),
Mewes et al. (1997), Andrews et al. (1997), Anderson and Kasevich (1998),
Bloch et al. (1999)]. It is also directly relevant to the He-in-vycor BEC
experiments [Crooker et al. (1983), Chan et al. (1988), Crowell et al.
(1995)]. The dynamics and statistics of the condensate is then obtained from
the canonical partition function. However the $N$-particle constraint
associated with the canonical ensemble is rather cumbersome and no simple
analytic expressions for the canonical partition function are known to exist
for three-dimensional traps. Even numerical calculations for large $N$ may
become impractical. A way out is to calculate the grand canonical properties
for the ideal Bose gas where the constraint of fixed particle number is
relaxed. This was how Einstein derived the characteristics of the condensate
and obtained the expression for the critical temperature. In general, we
would expect that the macroscopic properties of the condensate for both
canonical and grand canonical ensembles should be equivalent. However, as we
discuss below, only properties related to mean number of condensed particles
are almost identical in the two ensembles and the mean-square fluctuations
are remarkably different. Even as the temperature $T$ approaches zero when
all $N $ particles condense in the ground state, the fluctuations in the
grand canonical ensemble becomes huge, of the order of $N^2$, as discussed
below. This is clearly unacceptable.

Recently, realizing the inherent similarity in the phase transition behavior
between laser and the Bose-Einstein condensation, a new approach is
developed to study the nonequilibrium approach to BEC in the canonical
ensemble using the methods employed in the quantum theory of laser [Scully
(1999), Kocharovsky et al. (2000a)]. The advantage of this approach is that
analytic, though approximate, expressions are obtained for the canonical
partition function for the Bose-Einstein condensate for arbitrary traps. The
various moments for the condensate based on these analytic expressions and
the exact numerical results are in most cases negligible. This approach also
allows us to extend the critical temperature concept to the mesoscopic
systems, involving say $10^3$ atoms, in a natural fashion.

However, before proceeding to give the details of the laser theory based
analysis of BEC, we recall Einstein's arguments based on grand canonical
ensemble and see whether we can extend these arguments in a natural way to
describe a mesoscopic system. We also present the salient features of the
quantum theory of laser that become relevant in seeing the close connection
between the noneqilibrium approach to the dynamics and statistics of the
condensate of $N$-atom Bose gas and the photons inside a laser.

\subsection{Average condensate particle number}

Here we present a derivation of the average condensate particle number
following the original derivation of Einstein. We recall that Einstein
considered particles inside a box in the thermodynamic limit. We consider
particles in a harmonic trap and first discuss the thermodynamic limit. The
difference between a box and the harmonic trap is in the density of states.
We then go on to consider the mesoscopic number of particles.

Following Einstein we work with the grand canonical ensemble in which the
average condensate particle number $\bar{n}_{0}$ is determined as follows
[Ketterle and Druten (1996)]. The total number of atoms $N$ in the trap is
given by%
\begin{equation}
N=\sum_{k=0}^{\infty }\bar{n}_{\mathbf{k}}=\sum_{\mathbf{k}=0}^{\infty }%
\frac{1}{\exp [\beta (\varepsilon _{\mathbf{k}}-\mu )]-1},  \label{c1x}
\end{equation}%
where for the three dimensional (3D) isotropic harmonic trap we have $%
\varepsilon _{\mathbf{k}}=\hbar \Omega (k_{x}+k_{y}+k_{z})$, $\Omega $ is
the trap frequency, $\beta =1/k_{B}T$ and $\mu $ is the chemical potential.

In the following we demonstrate how to calculate the mean number of
condensed particles $\bar{n}_{0}$ for a 3D isotropic harmonic trap. Using $%
\bar{n}_{0}=1/(exp(-\beta \mu )-1)$, we can relate the chemical potential $%
\mu $ to $\bar{n}_{0}$ as $1+1/\bar{n}_{0}=\exp (-\beta \mu )$ and rewrite
Eq. (\ref{c1x}) as 
\begin{equation}
N=\sum_{\mathbf{k}=0}^{\infty }\langle n_{\mathbf{k}}\rangle =\sum_{\mathbf{k%
}=0}^{\infty }\frac{1}{(1+1/\bar{n}_{0})\exp (\beta \varepsilon _{\mathbf{k}%
})-1}.  \label{c2x}
\end{equation}%
For large $\bar{n}_{0}$ we neglect $1/\bar{n}_{0}$ in copmparison with 1.
Following Einstein, we proceed to separate off the ground state so that Eq. (%
\ref{c2x}) can be written as 
\begin{equation}
N-\bar{n}_{0}=\mathcal{H}\;,  \label{sum1}
\end{equation}%
where%
\begin{equation}
\mathcal{H}=\sum_{k>0}\frac{1}{e^{\beta \epsilon _{k}}-1}.  \label{c3x}
\end{equation}%
For an isotropic harmonic trap with frequency $\Omega $ the degeneracy of
the $n$th energy level is $(n+2)(n+1)/2$, and we obtain 
\begin{equation}
\mathcal{H}=\frac{1}{2}\sum_{n=1}^{\infty }\frac{(n+2)(n+1)}{\exp (\beta
n\hslash \Omega )-1}\approx \frac{1}{2}\int_{1}^{\infty }\frac{(x+2)(x+1)}{%
\exp (x\beta \hslash \Omega )-1}dx.  \label{sum2}
\end{equation}%
In the limit $k_{B}T\gg \hslash \Omega $ we find 
\begin{equation}
\mathcal{H}\approx \frac{1}{2}\int_{0}^{\infty }\frac{x^{2}}{\exp (x\beta
\hslash \Omega )-1}dx=\left( \frac{k_{B}T}{\hslash \Omega }\right) ^{3}\zeta
(3),  \label{sum3}
\end{equation}%
where $\zeta (x)$ is the Riemann zeta-function. We define the critical
temperature $T_{c}$ such that when $T=T_{c}$ we have $\bar{n}_{0}=0$. This
yields 
\begin{equation}
T_{c}=\frac{\hslash \Omega }{k_{B}}\left( \frac{N}{\zeta (3)}\right) ^{1/3}
\label{critt}
\end{equation}%
as the temperature of BEC transition in the thermodynamic limit. The
resulting expression for the mean number of particles in the condensate is 
\begin{equation}
\bar{n}_{0}(T)=N\left[ 1-\left( \frac{T}{T_{c}}\right) ^{3}\right]
\label{crit}
\end{equation}%
which shows a cusp at $T=T_{c}$. For mesoscopic number of particles (e.g., a
few hundred) Eq. (\ref{crit}) becomes inaccurate as the thermodynamic limit
is not reached. To improve the accuracy we first rewrite Eq. (\ref{c2x}) in
the following way [Kocharovsky et al. (2006), Jordan et al. (2006)]

\begin{equation}
N-\bar{n}_{0}=\frac{1}{\left( \frac{1}{\bar{n}_{0}}+1\right) }%
\sum\limits_{k>0}\frac{1}{e^{\beta \epsilon _{k}}-\frac{\bar{n}_{0}}{\bar{n}%
_{0}+1}}.  \label{c4x}
\end{equation}%
For $\bar{n}_{0}\gg 1$, the term $\bar{n}_{0}/(\bar{n}_{0}+1)$ inside the
summation may be approximated by $1$. Then we obtain a quadratic equation
for the mean number of particles in the ground state 
\begin{equation}
N-\bar{n}_{0}=\frac{\mathcal{H}}{\frac{1}{\bar{n}_{0}}+1}\qquad
\Longrightarrow \qquad \bar{n}_{0}^{2}+\bar{n}_{0}(1+\mathcal{H}-N)-N=0
\end{equation}%
whose solution is 
\begin{equation}
\bar{n}_{0}=\frac{1}{2}\left( N-\mathcal{H}-1+\sqrt{(N-\mathcal{H}-1)^{2}+4N}%
\right) .  \label{c5x}
\end{equation}

\begin{figure}[h]
\begin{center}
\includegraphics[width=8cm]{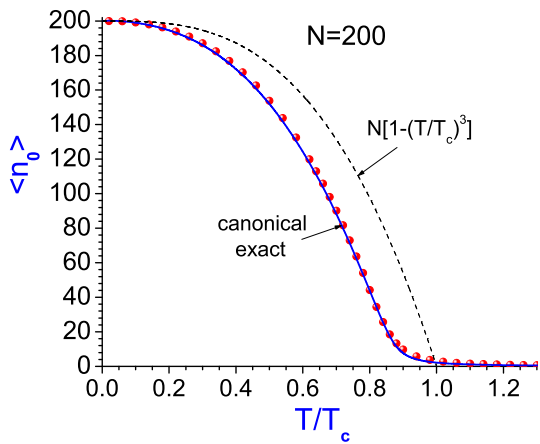}
\end{center}
\caption{ The average condensate particle number versus temperature for $%
N=200 $ particles in an isotropic harmonic trap. Solid line is Eq. (\protect
\ref{c5x}), while the dash line shows the thermodynamic limit formula (%
\protect\ref{crit}). ``Exact" dots are obtained numerically in the canonical
ensemble [Wilkens and Weiss (1997)].}
\label{n0}
\end{figure}

Analytical expression (\ref{c5x}) shows a smooth crossover near $T_{c}$ for
a mesoscopic number of particles $N$ as shown in Fig. \ref{n0}. Here we
compare the mean condensate number as given by Eq.~(\ref{c5x}) obtained in
the grand canonical ensemble (solid line) for $N=200$ and the solution (\ref%
{crit}) (dashed line) that is valid only for a large number of particles $N$
with the numerical calculation of $\bar{n}_{0}(T)$ from the exact recursion
relations in the canonical ensemble (dots) [Wilkens and Weiss (1997)]. In
the canonical ensemble the total number of particles $N$ is fixed, rather
then the chemical potential. We see that, for the average particle number,
both ensembles (grand canonical and canonical) yield very close answers. The
interesting observation is that the approximate expression (\ref{crit})
obtained in a suitable limit within the grand canonical ensemble yields
results that are indistinguishable from the exact results from the canonical
ensemble. However, as we discuss be low, this is not the case for the BEC
fluctuations.

%\begin{figure}[h]
%\center\epsfxsize=8.0cm\epsffile{r1.eps}
%\caption{The average condensate particle number versus temperature for $%
%N=200 $ particles in an isotropic harmonic trap. Solid line is Eq. (\protect
%\ref{c5x}), while the dash line shows the thermodynamic limit formula (%
%\protect\ref{crit}). ``Exact" dots are obtained numerically in the canonical
%ensemble [Wilkens and Weiss (1997)].}
%\label{n0}
%\end{figure}

\subsection{Fluctuations in the number of particles in the condensate}

Condensate fluctuations are characterized by the central moments $%
\mu_{m}=\langle (n_{0}-\bar{n}_{0})^{m}\rangle $. The first of them is the
squared variance 
\begin{equation}
\langle (n_{0}-\bar{n}_{0})^{2}\rangle =\langle n_{0}^{2}\rangle -\langle
n_{0}\rangle ^{2},
\end{equation}

When the temperature~$T$ approaches zero, all $N$~particles are forced into
the system's ground state, so that the mean square $\langle (\Delta
n_{0})^{2}\rangle $ of the fluctuation of the ground-state occupation number
has to vanish for $T\rightarrow 0$. However the grand canonical description
gives $\langle (\Delta n_{0})^{2}\rangle \rightarrow N(N+1)$, clearly
indicating that with respect to these fluctuations the different statistical
ensembles are no longer equivalent. What, then, would be the correct
expression for the fluctuation of the ground-state occupation number within
the \emph{canonical\/} ensemble, which excludes any exchange of particles
with the environment, but still allows for the exchange of energy? Various
aspects of this riddle have appeared in the literature over the years [Ziff
et al. (1977), ter Haar (1970), Fierz (1956)], mainly inspired by academic
curiosity, before it resurfaced in 1996 [Grossmann and Holthaus (1996),
Politzer (1996), Gajda and Rz\c{a}\.{z}ewski (1977), Wilkens and Weiss
(1997), Weiss and Wilkens (1997)], this time triggered by the experimental
realization of mesoscopic Bose--Einstein condensates in isolated micro
traps. Condensate fluctuations can be measured by means of a scattering of
series of short laser pulses [Idziaszek (2000)] (see also [Chuu et al.
(2005)]. Since then, much insight into this surprisingly rich problem has
been gained. Much of this insight follows directly from the quantum theory
of the laser, to which we now turn.

\section{The Quantum Theory of the Laser}

\setcounter{equation}{0}

The quantum (photon) picture of maser/laser operation is a difficult problem
in the interaction of radiation with matter. Even several years after the
development of the maser and the laser there was not a fully quantized
theory of laser action. The difficulties inherent in this problem were most
succinctly stated by Roy Glauber in his 1964 Les Houches lectures in this
way [Glauber (1964)]:

\textquotedblleft \textit{The only reliable method we have of constructing
density operators, in general, is to devise theoretical models of the system
under study and to integrate [the] corresponding Schr\"{o}dinger equation,
or equivalently to solve the equation of motion for the density operator.
These assignments are formidable ones for the case of the laser oscillator
and have not been carried out to date in quantum mechanical terms. The
greatest part of the difficulty lies in the mathematical complications
associated with the nonlinearity of the device. The nonlinearity physics
plays an essential role in stabilizing the field generated by the laser. It
seems unlikely, therefore, that we shall have a quantum mechanically
consistent picture of the frequency bandwidth of the laser or of the
fluctuations of its output until further progress is made with these
problems.}"

Following the Les Houches meeting, Marlan Scully and Willis Lamb took up the
challenge and developed a fully quantum mechanical theory of laser that
yielded the photon statistics above, at, and below threshold (diagonal
elements of the laser field density matrix), showed that the laser linewidth
was contained in the time decay of the off-diagonal elements of the density
matrix, and made the physics clearer by comparing the laser threshold to a
ferromagnetic phase transition [Scully and Lamb (1966)]. They presented
their theory of ``optical maser" at the famous Puerto Rico Conference on the
``Physics of Quantum Electronics" in the summer of 1965.

The treatment of the laser near threshold must include nonlinear active
medium consisting of atoms that are pumped in their excited states and a
damping mechanism to account for the loss of photons from the cavity through
end mirrors. To obtain laser pumping action we introduce atoms in their
upper level $|a>$ at random times $t_{i}$, decaying to a far-removed ground
state $|b>$. Cavity field damping is included by coupling the field to an
ensemble of atoms in their ground state ($\gamma $ subsystem in Fig. \ref%
{las1}).

Here we concentrate on the study of the photon distribution function for the
laser field which is given by the diagonal matrix elements of the reduced
density operator of the field. The photon statistical distribution for the
laser is of interest for several reasons. Historically, it was initially
thought by some that the statistical photon distribution should be a
Bose-Einstein distribution. A little reflection shows that this can not be,
since the laser is operating far from thermodynamic equilibrium. However, a
different paradigm recognizes many atoms oscillating in phase produce what
is essentially a classical current, and this would generate a coherent
state; the statistics of which is Possionian. But, for example, the photon
statistics of a typical Helium-Neon laser is substantially different from a
Possionian distribution. Of course, well above threshold, the steady-state
laser photon statistical distribution is Poisson which is the characteristic
of a coherent state. In order to see these interesting features we consider
the master equation of the laser in various regimes of operation.

\begin{figure}[h]
\begin{center}
\includegraphics[width=11cm]{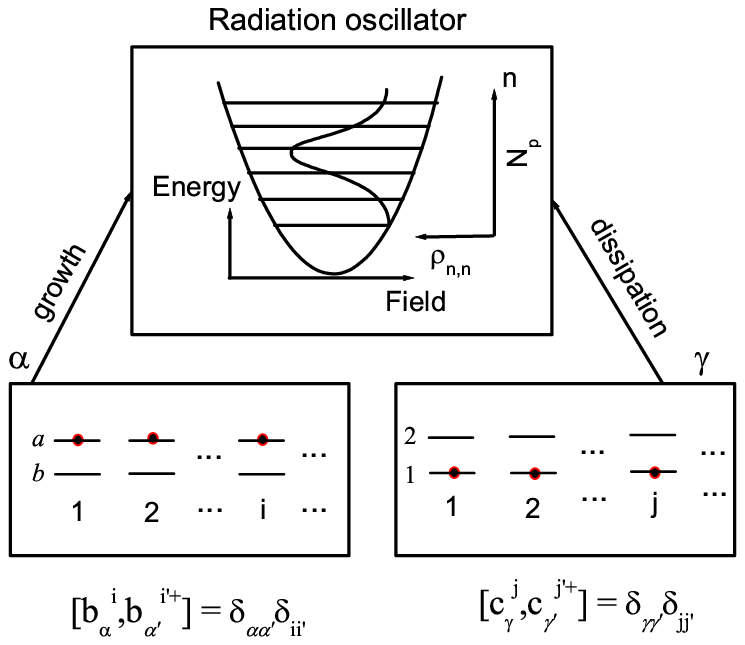}
\end{center}
\caption{Model.}
\label{las1}
\end{figure}

%\begin{figure}[h]
%\center\epsfxsize=11.0cm\epsffile{model.eps}
%\caption{Model.}
%\label{las1}
%\end{figure}

Here we omit details of the full theory and point out that the diagonal
elements $\rho _{nn}\equiv p(n)$, which represent the probability of $n$
photons in the field, satisfy the equation of motion%
\[
\dot{p}(n)=-\left[ {\frac{(n+1){\mathcal{A}}}{1+(n+1){\mathcal{B}}/\mathcal{A%
})}}\right] p(n)+\left( {\frac{n{\mathcal{A}}}{1+n{\mathcal{B}}/{\mathcal{A}}%
}}\right) p(n-1) 
\]%
\begin{equation}
-{\mathcal{C}}np(n)+{\mathcal{C}}(n+1)p(n+1).  \label{lasermaster}
\end{equation}
where $\mathcal{A}$ is the linear gain coefficient, $\mathcal{B}$ is the
self-saturation coefficient and $\mathcal{C}$ is the decay rate. The first
two terms in the right hand side of Eq. (\ref{lasermaster}) describe pumping
and the last two terms come from damping (decay). It is interesting to note
that the diagonal elements are coupled only to diagonal elements. More
generally, only off-diagonal elements $\rho _{nn^{\prime }}$ with the same
difference $(n-n^{\prime })$ are coupled.

\begin{figure}[h]
\begin{center}
\includegraphics[width=8cm]{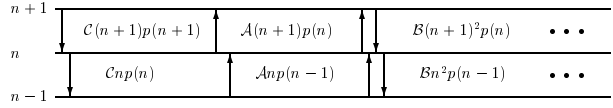}
\end{center}
\caption[Detailed balance and the corresponding probability flow diagram.]{
Detailed balance and the corresponding probability flow diagram.}
\label{las2}
\end{figure}

Before we begin the solution of the above equation we want to give a simple
intuitive physical picture of the processes it describes in terms of a
probability flow diagram, Fig.\ \ref{las2}.

The left-hand-side is the rate of change of the probability of finding $n$
photons in the cavity. The right-hand-side contains the physical processes
that contribute to the change. Each process is represented by an arrow in
the diagram. The processes are proportional to the probability of the state
they are starting from and this will be the starting point of the arrow. The
tip of the arrow points to the state the process is leading to.

%\begin{figure}[h]
%\center\epsfxsize=8.0cm\epsffile{schem.eps}
%\caption{Detailed balance and the corresponding probability flow diagram.}
%\label{las2}
%\end{figure}

A simple physical meaning can be given to Eq.\ (\ref{lasermaster}) for the
photon distribution function in terms of a probability flow diagram (Fig.\ %
\ref{las2}) by expanding the terms in the denominator of Eq.\ (\ref%
{lasermaster}). There we see the `flow' of probability in and out of the $%
|n\rangle$ state from and to the neighboring $|n+1\rangle$ and $|n-1\rangle$
states. For example, the ${\mathcal{A}}(n+1) p(n)$ term represents the flow
of probability from the $|n\rangle$ state to the $|n+1\rangle$ state due to
the emission of photons by lasing atoms initially in the upper states. Here $%
{\mathcal{A}}n$ is the rate of stimulated emission, ${\mathcal{A}}$ is the
spontaneous emission rate and these rates are multiplied by $p(n)$ to yield
the total probability flow rate. Since the probability flows out of $p(n)$,
this term is negative. The first term in the expansion of the
square-bracketed term in (\ref{lasermaster}), namely ${\mathcal{B}}(n+1)^2
p(n) = \mathcal{A} (n+1) ({\mathcal{B}}/{\mathcal{A}}) (n+1) p(n)$,
corresponds to the process in which photons are emitted and then reabsorbed,
the reabsorption rate being $(\mathcal{B}/{\mathcal{A}}) (n+1)$. Similar
explanations exist for the other terms including the loss terms.

After this brief discussion of the meaning of the individual terms we now
turn our attention to the solution of the laser master equation (\ref%
{lasermaster}). Although it is possible to obtain a rather general
time-dependent solution to Eq. (\ref{lasermaster}), our main interest here
is in the steady-state properties of the field. To obtain the steady-state
photon statistics, we replace the time derivative with zero. Notice that the
right-hand-side of the equation is of the form $F(n+1)-F(n)$, where 
\begin{equation}
F(n)=\mathcal{C}np(n)-\frac{n\mathcal{A}}{1+n\mathcal{B}/\mathcal{A}}p(n-1),
\label{Fdef}
\end{equation}%
simply meaning that in steady-state $F(n+1)=F(n)$. In other words $F(n)$ is
independent of $n$ and is, therefore, a constant $c$. Furthermore, the
equation $F(n)=c$ has normalizable solution only for $c=0$. From Eq. (\ref%
{Fdef}) we then immediately obtain 
\begin{equation}
p(n)=\frac{\mathcal{A}/\mathcal{C}}{1+n\mathcal{B}/\mathcal{A}}p(n-1),
\label{lasrec}
\end{equation}%
which is a very simple two-term recurrence relation to determine the
photon-number distribution. Before we present the solution a remark is
called for here. The fact that $F(n)=0$ and $F(n+1)=0$ hold separately is
called the condition of detailed balance. As a consequence we do not need to
deal with all four processes affecting $p(n)$. It is sufficient to balance
the processes connecting a pair of adjacent levels in Fig. \ref{las2} and
instead of solving the general three-term recurrence relation, resulting
from the steady state version of Eq. (\ref{lasermaster}), it is enough to
solve the much simpler two-term recursion, Eq. (\ref{lasrec}).

It is instructive to investigate the photon statistics in some limiting
cases before discussing the general solution. Below threshold the linear
approximation holds. Since only very small $n$ states are populated
appreciably, the denominator on the right-hand-side of (\ref{lasrec}) can be
replaced by unity in view of $n\mathcal{B}/\mathcal{A}\ll 1$. Then 
\begin{equation}
p(n)=p(0)\left( \frac{\mathcal{A}}{\mathcal{C}}\right) ^{n}.
\end{equation}%
The normalization condition, $\sum_{n=0}^{\infty }p(n)=1$, determines the
constant $p(0)$, yielding $p(0)=(1-\mathcal{A}/\mathcal{C})$. Finally 
\begin{equation}
p(n)=(1-\mathcal{A}/\mathcal{C})\left( \frac{\mathcal{A}}{\mathcal{C}}%
\right) ^{n}.  \label{lphotstat1}
\end{equation}%
Clearly, the condition of existence for this type of solution is $\mathcal{A}%
<\mathcal{C}$. Therefore, $\mathcal{A}=\mathcal{C}$ is the threshold
conditon for the laser. At threshold, the photon statistics changes
qualitatively and very rapidly in a narrow region of the pumping parameter.
It should also be noted that below threshold the distribution function (\ref%
{lphotstat1}) is essentially of thermal character. If we introduce an
effective temperature $T$ defined by 
\begin{equation}
e^{-\hbar \omega _{0}/kT}=\mathcal{A}/\mathcal{C},
\end{equation}%
we can cast (\ref{lphotstat1}) to the form 
\begin{equation}
p(n)=(1-e^{-\hbar \omega _{0}/kT})e^{-n\hbar \omega _{0}/kT}.  \label{c16}
\end{equation}%
This is just the photon number distribution of a single mode in thermal
equilibrium with a thermal reservoir at temperature $T$. The inclusion of a
finite temperature loss reservoir to represent cavity losses will not alter
this conclusion about the region below threshold.

There is no real good analytical approximation for the region around
threshold although the lowest order expansion of the denominator in (\ref%
{lasrec}) yields some insight. The solution with this condition is given by 
\begin{equation}
p(n) = p(0) \left( \frac{\mathcal{A}}{\mathcal{C}}\right)^{n}
\prod_{k=0}^{n} (1-k\mathcal{B}/\mathcal{A}) .  \label{c13}
\end{equation}
This equation clearly breaks down for $n>\mathcal{A}/\mathcal{B} =n_{\mathrm{%
max}}$, where $p(n)$ becomes negative. The resulting distribution is quite
broad exhibiting a long plateau and a rapid cut-off at $n_{\mathrm{max}}$.
The broad plateau means that many values of $n$ are approximately equally
likely and, therefore, the intensity fluctuations are large around
threshold. The most likely value of $n=n_{\mathrm{opt}}$ can be obtained
from the condition $p(n_{\mathrm{opt}}-1)=p(n_{\mathrm{opt}})$ since $p(n)$
is increasing before $n=n_{\mathrm{opt}}$ and decreasing afterward. This
condition yields $n_{\mathrm{opt}}=(\mathcal{A} -\mathcal{C})/\mathcal{B}$
which is smaller by the factor $\mathcal{C} / \mathcal{A}$ than the value
obtained from the full nonlinear equation.

The third region of special interest is the one far above threshold. In this
region $\mathcal{A}/\mathcal{C}\gg 1$ and the $n$ values contributing the
most to the distribution function are the ones for which $n\gg \mathcal{A}/%
\mathcal{B}$. We can then neglect $1$ in the denominator of (\ref{lasrec}),
yielding 
\begin{equation}
p(n)=e^{-\bar{n}}\frac{{\bar{n}}^{n}}{n!},  \label{Poisson}
\end{equation}%
with ${\bar{n}}={\mathcal{A}}^{2}/(\mathcal{C}\mathcal{B})$. Thus the photon
statistics far above threshold are Poissonian, the same as for a coherent
state. This, however, does not mean that far above threshold the laser is in
a coherent state. As we shall see later, the off-diagonal elements of the
density matrix remain different from those of a coherent state for all
regimes of operation.

\begin{figure}[h]
\begin{center}
\includegraphics[width=9cm]{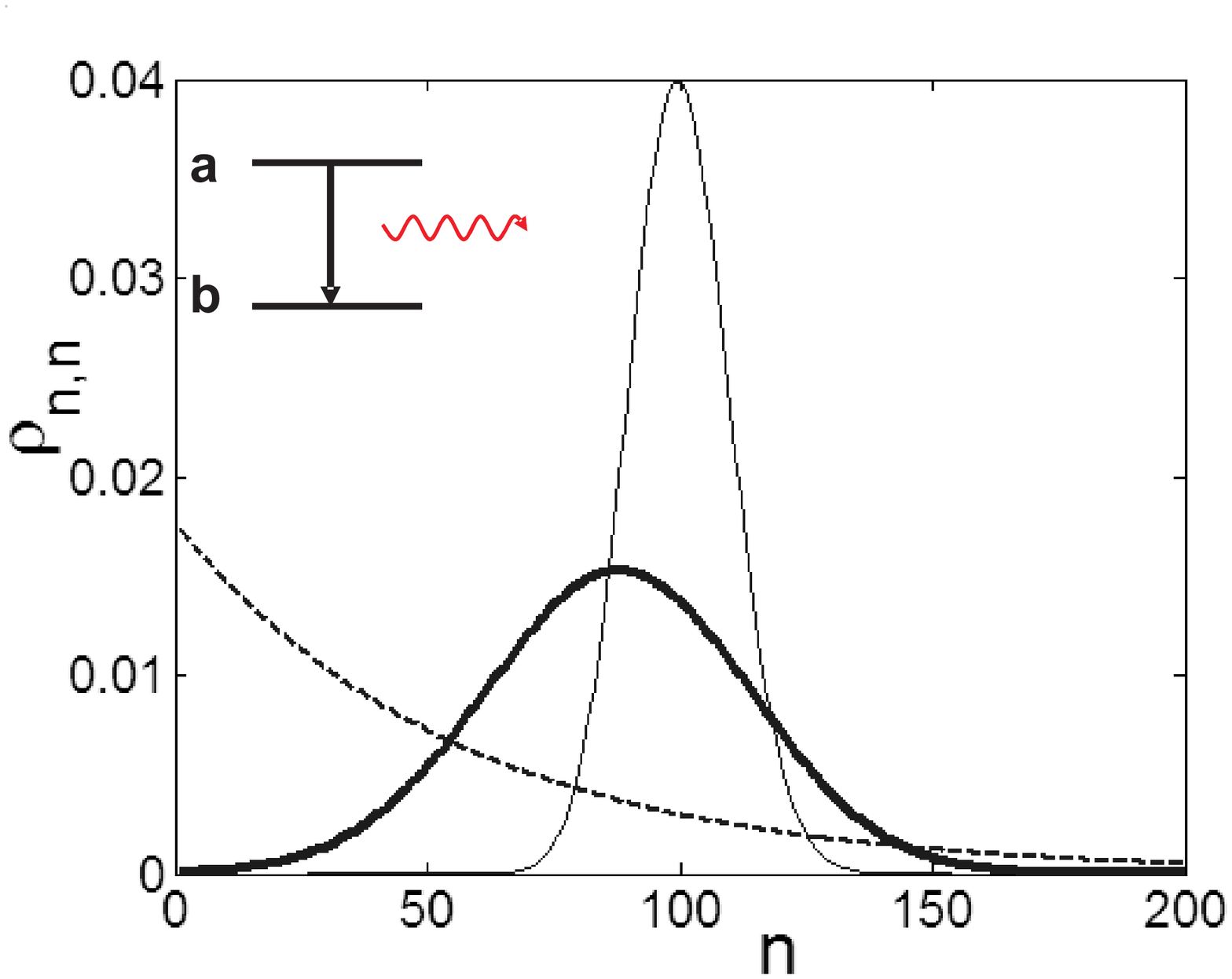}
\end{center}
\caption{Photon number distributions for a) thermal photons plotted from Eq.
(\protect\ref{c16}) (dashed line), b) coherent state (Poissonian) (thin
solid line), and c) He-Ne laser plotted using Eq. (\protect\ref{c13}) (thick
solid line). Insert shows an atom making a radiation transition.}
\label{las3}
\end{figure}

Figure \ref{las3} shows photon number distribution in different limits.

%\begin{figure}[tbp]
%\center \epsfxsize=9cm\epsffile{tl.eps}
%\caption{Photon number distributions for a) thermal photons plotted from Eq.
%(\protect\ref{c16}) (dashed line), b) coherent state (Poissonian) (thin
%solid line), and c) He-Ne laser plotted using Eq. (\protect\ref{c13}) (thick
%solid line). Insert shows an atom making a radiation transition.}
%\label{las3}
%\end{figure}

\section{Bose-Einstein Condensation: Laser Phase-Transition Analogy}

\setcounter{equation}{0}

Bose-Einstein condensation in a trap has intriguing similarities with the
threshold behavior of a laser which also can be viewed as a kind of a phase
transition [DeGiorgio and Scully (1970), Graham and Haken (1970),
Kocharovsky et. al. (2006)]. In both cases stimulated processes are
responsible for the appearance of the macroscopic order parameter. The main
difference is that for the Bose gas in a trap there is also interaction
between the atoms which, in particular, yields stimulated effects in BEC. On
the other hand there are two subsystems for the laser, namely the laser
field and the active atomic medium. The crucial point for lasing is the
interaction between the field and the atomic medium. Thus, the effects of
different interactions in the laser are easy to trace and relate to the
observable characteristics of the system. This is not the case in BEC and it
is important to separate different effects.

As we discussed in the previous section, the laser light is conveniently
described by a master equation obtained by treating the atomic (gain) media
and cavity dissipation (loss) as reservoirs which when \textquotedblleft
traced over" yield the coarse grained equation of motion for the reduced
density matrix for laser radiation. We thus arrive at the equation of motion
for the probability of having $n$ photons in the cavity given by Eq.~(\ref%
{lasermaster}). From Eq.~(\ref{c13}) we have that partially coherent laser
light has a sharp photon distribution (with width several times Poissonian
for a typical He-Ne laser) due to the presence of the saturation
nonlinearity, $\mathcal{B}$, in the laser master equation. Thus, we see that
the saturation nonlinearity in the radiation-matter interaction is essential
for laser coherence.

Next we turn to an ideal Bose gas and derive a master equation for the
particles in the condensate. The steady-state description of the condensate
arises from the inherent nonlinearities in the system. One naturally asks:
Is the corresponding nonlinearity in BEC due to atom-atom scattering? or Is
there a nonlinearity present even in an ideal Bose gas? In the following we
show that the latter is the case.

\subsection{Condensate master equation}

Here we consider a model of a dilute Bose gas of atoms wherein interatomic
scattering is neglected. This ideal Bose gas of $N$ atoms is confined inside
a trap and the atoms exchange energy with a reservoir at a fixed temperature 
$T$ [Scully (1999), Kocharovsky et al. (2000a), Kapale and Zubairy (2001)].
The ``ideal gas + reservoir\textquotedblright\ model corresponds to a
canonical ensemble and it allows us to demonstrate most clearly the master
equation approach to the analysis of dynamics and statistics of BEC. It
provides the simplest description of many qualitative and, in some cases,
quantitative characteristics of the experimental BEC. In particular, it
explains many features of the condensate dynamics and fluctuations and
allows us to obtain the particle number statistics of the BEC. An extension
of the present approach to the case of an interacting gas which includes
usual many-body effects due to interatomic scattering will be discussed in
the next section.

\begin{figure}[h]
\begin{center}
\includegraphics[width=10cm]{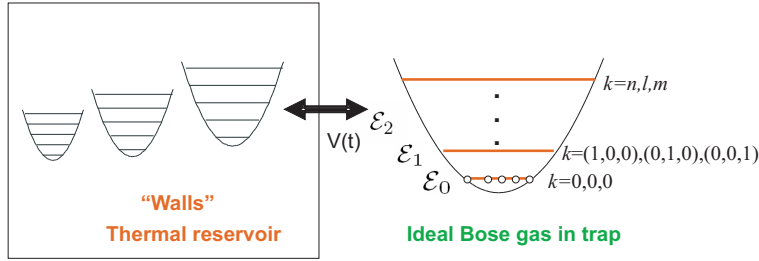}
\end{center}
\caption{Simple harmonic oscillators as a thermal reservoir for the ideal
Bose gas in a trap.}
\label{reservoir}
\end{figure}

For many problems a concrete realization of the reservoir system is not very
important if its energy spectrum is dense and flat enough. For example, one
expects (and we find) that the equilibrium (steady state) properties of the
BEC are largely independent of the details of the reservoir. For the sake of
simplicity, we assume that the reservoir is an ensemble of simple harmonic
oscillators whose spectrum is dense and smooth, see Fig. \ref{reservoir}.
The interaction between the gas and the reservoir is described by the
interaction picture Hamiltonian 
\begin{equation}
V=\sum_{j}\sum_{k>l}g_{j,kl}b_{j}^{\dagger }a_{k}a_{l}^{\dagger
}e^{-i(\omega _{j}-\nu _{k}+\nu _{l})t}+H.c.,  \label{II9}
\end{equation}%
where $b_{j}^{\dagger }$ is the creation operator for the reservoir $j$
oscillator (\textquotedblleft phonon"), and $a_{k}^{\dagger }$ and $a_{k}$ ($%
k\neq 0$) are the creation and annihilation operators for the Bose gas atoms
in the $k\mathrm{th}$ level. Here $\hbar \nu _{k}$ is the energy of the $k%
\mathrm{th}$ level of the trap, and $g_{j,kl}$ is the coupling strength.

%\begin{figure}[tbp]
%\center\epsfxsize=10cm\epsffile{reservoir.eps}
%\caption{Simple harmonic oscillators as a thermal reservoir for the ideal
%Bose gas in a trap.}
%\label{reservoir}
%\end{figure}

Following along the lines of the quantum theory of the laser we can derive
an equation of motion for the distribution function of the condensed bosons (%
$p_{n_{0}}\equiv \rho _{n_{0},n_{0}}$) [Kocharovsky et al. (2000a)]%
\[
\dot{p}_{n_{0}}=-\kappa
\{K_{n_{0}}(n_{0}+1)p_{n_{0}}-K_{n_{0}-1}n_{0}p_{n_{0}-1} 
\]%
\begin{equation}
+H_{n_{0}}n_{0}p_{n_{0}}-H_{n_{0}+1}(n_{0}+1)p_{n_{0}+1}\},  \label{II20}
\end{equation}%
where $\kappa $ embody the spectral density of the bath and the coupling
strength of the bath oscillators to the gas particles and 
\begin{equation}
K_{n_{0}}=\sum_{k>0}(\eta _{k}+1)\langle n_{k}\rangle _{n_{0}},\qquad
H_{n_{0}}=\sum_{k>0}\eta _{k}(\langle n_{k}\rangle _{n_{0}}+1),  \label{II21}
\end{equation}%
with 
\begin{equation}
\eta _{k}=\frac{1}{e^{\hbar \nu _{k}/T}-1}.
\end{equation}%
The particle number constraint comes in since $\sum_{k>0}\langle
n_{k}\rangle _{n_{0}}=N-\bar{n}_{0}$.

The steady state distribution of the number of atoms condensed in the ground
level of the trap can be determined from Eq. (\ref{II20}) and the various
moments, including the mean value and the variance, can then be determined.
It is clear that there are two processes: cooling and heating. The cooling
process is represented by the first two terms with the cooling coefficient $%
K_{n_{0}}$, and the heating by the third and fourth terms with heating
coefficient $H_{n_{0}}$. In the cooling process the atoms in the excited
atomic levels in the trap jump to the condensate level and transfer energy
to the thermal reservoir whereas, in the heating process, the atoms in the
condensate absorb energy from the reservoir and get excited. The cooling and
heating coefficients have the analogy with the saturated gain and cavity
loss in the laser master equation (\ref{lasermaster}). According to Eq. (\ref%
{II21}), these coefficients depend upon trap parameters such as the shape of
the trap, the total number of bosons in the trap, $N$, and the temperature $%
T $.

In general, the cooling and the heating coefficients are complicated and
depend upon the condensate probability distribution $p_{n_0}$. In this
sense, Eq. (\ref{II20}) is a transcendental equation for $p_{n_0}$. This
equation can however be simplified in certain approximations and we obtain
analytic results for the condensate distribution that are close to the exact
numerical results.

\subsection{Low temperature approximation}

At low enough temperatures, the average occupations in the reservoir are
small and $\eta _{k}+1\simeq 1$ in Eq. (\ref{II21}). This suggests the
simplest approximation for the cooling coefficient 
\begin{equation}
K_{n_{0}}\simeq \sum_{k>0}\langle n_{k}\rangle _{n_{0}}=N-\bar{n}_{0}.
\label{II26}
\end{equation}%
In addition, at very low temperatures the number of non-condensed atoms is
also very small. We can therefore approximate $\langle
n_{k}\rangle_{n_{0}}+1 $ by $1$ in Eq. (\ref{II21}). Then the heating
coefficient is a constant equal to the total average number of thermal
excitations in the reservoir at all energies corresponding to the energy
levels of the trap, 
\begin{equation}
H_{n_{0}}\simeq \mathcal{H},\qquad \mathcal{H}\equiv \sum_{k>0}\eta
_{k}=\sum_{k>0}\frac{1}{\left( e^{\hbar \nu _{k}/T}-1\right) }.  \label{II27}
\end{equation}

Under these approximations, the condensate master equation (\ref{II20})
simplifies considerably and contains only one non-trivial parameter $%
\mathcal{H}$. We obtain%
\[
\dot{p}_{n_{0}}=-\kappa
\{(N-n_{0})(n_{0}+1)p_{n_{0}}-(N-n_{0}+1)n_{0}p_{n_{0}-1} 
\]
\begin{equation}
+\mathcal{H}[n_{0}p_{n_{0}}-(n_{0}+1)p_{n_{0}+1}]\}.  \label{II28}
\end{equation}%
It may be noted that Eq. (\ref{II28}) has the same form as the equation~of
motion for the photon distribution function in a laser operating not too far
above threshold ($\mathcal{B}\langle n\rangle /\mathcal{A}<<1)$). The
identification is complete if we define the gain, saturation, and loss
parameters in laser master equation by $\mathcal{A}=\kappa (N+1)$, $\mathcal{%
B}=\kappa $, and $\mathcal{C}=\kappa \mathcal{H}$, respectively. The
mechanism for gain, saturation, and loss are however different in the
present case.

The resulting steady state distribution for the number of condensed atoms is
given by 
\begin{equation}
p_{n_0}=\frac {1}{Z_N}\frac{\mathcal{H}^{N-n_0}}{(N-n_0)!},  \label{II30}
\end{equation}
where $Z_N=1/p_N$ is the partition function. It follows from the
normalization condition $\sum_{n_0}p_{n_0}=1$ that 
\begin{equation}
Z_N={e^{\mathcal{H}}}\Gamma(N+1,\mathcal{H})/N!,  \label{II31}
\end{equation}
where $\Gamma (\alpha,x) = \int^{\infty}_{x} t^{\alpha -1} e^{-t} dt$ is an
incomplete gamma-function.

The mean value and the variance can be calculated from the distribution (\ref%
{II30}) for an arbitrary finite number of atoms in the Bose gas, 
\begin{equation}
\langle n_{0}\rangle =N-\mathcal{H}+\mathcal{H}^{N+1}/Z_{N}N!,  \label{II33}
\end{equation}
\begin{equation}
\Delta n_{0}^{2}\equiv \langle n_{0}^{2}\rangle -\langle n_{0}\rangle ^{2}=%
\mathcal{H}\left( 1-(\langle n_{0}\rangle +1)\mathcal{H}^{N}/Z_{N}N!\right) .
\label{II34}
\end{equation}

As we shall see from the extended treatment in the next section, the
approximations (\ref{II26}), (\ref{II27}) and, therefore, the results (\ref%
{II33}), (\ref{II34}) are clearly valid at low temperatures, i.e., in the
weak trap limit, $T\ll \varepsilon _{1}$, where $\varepsilon _{1}$ is an
energy gap between the first excited and the ground levels of a
single-particle spectrum in the trap. However, in the case of a harmonic
trap the results (\ref{II33}), (\ref{II34}) show qualitatively correct
behavior for all temperatures, including $T \gg \varepsilon _{1}$ and $T\sim
T_{c}$ [Scully (1999)].

In particular, for a harmonic trap we have from Eq. (\ref{II27}) that the
heating rate is 
\begin{equation}
\mathcal{H}=\sum_{l,m,n}\frac{1}{\exp [\beta \hbar \Omega (l+m+n)-1]}\approx
\left( \frac{k_{B}T}{\hbar \Omega }\right) ^{3}\zeta (3)=N\left( \frac{T}{%
T_{c}}\right) ^{3}.  \label{Hn0 harmonic}
\end{equation}%
Thus, in the low temperature region, the master equation~(\ref{II28}) for
the condensate in the harmonic trap becomes [Scully (1999)] 
\[
\frac{1}{\kappa }\dot{p}_{n_{0}}=-\left[ (N+1)(n_{0}+1)-(n_{0}+1)^{2}\right]
p_{n_{0}}+[(N+1)n_{0}-n_{0}^{2}]p_{n_{0}-1} 
\]%
\begin{equation}
-N\left( \frac{T}{T_{c}}\right) ^{3}[n_{0}p_{n_{0}}-(n_{0}+1)p_{n_{0}+1}].
\end{equation}
The resulting plots for $\bar{n}_{0}$, the variance and the third and fourth
central moments are given in Fig. \ref{m1234} (dash line). These analytical
results give qualitatively correct description of the ideal Bose gas when
compared with the exact solution for the moments as derived in the canonical
ensemble. These quantitative agreement with the exact numerical results can
be considerably improved in the quasithermal approximation to which we turn
next.

\subsection{Quasithermal approximation for non-condensate occupations}

At arbitrary temperatures, a very reasonable approximation for the average
non-condensate occupation numbers in the cooling and heating coefficients in
Eq. (\ref{II21}) is given by 
\begin{equation}
\langle n_{k}\rangle _{n_{0}}=\eta _{k}\sum_{k>0}\langle n_{k}\rangle
_{n_{0}}/\sum_{k^{\prime }>0}\eta _{k^{\prime }}=\frac{(N-\bar{n}_{0})}{%
(e^{\varepsilon _{k}/T}-1){\mathcal{H}}}\;,  \label{II35}
\end{equation}%
Equation (\ref{II35}) satisfies the canonical-ensemble constraint, $%
N=n_{0}+\sum_{k>0}n_{k}$, independently of the resulting distribution $%
p_{n_{0}}$. This important property is based on the fact that a quasithermal
distribution (\ref{II35}) provides the same relative average occupations in
excited levels of the trap as in the thermal reservoir.

The cooling and heating coefficients (\ref{II21}) in the quasithermal
approximation of Eq. (\ref{II35}) are 
\begin{equation}
K_{n_{0}}=(N-n_{0})(1+\eta ),\qquad H_{n_{0}}=\mathcal{H}+(N-n_{0})\eta .
\label{II36}
\end{equation}%
Compared with the low temperature approximation (\ref{II26}) and (\ref{II27}%
), these coefficients acquire an additional contribution $(N-n_{0})\eta $
due to the cross-excitation parameter, i.e., 
\begin{equation}
\eta =\frac{1}{N-n_{0}}\sum_{k>0}\langle \eta _{k}\rangle \langle
n_{k}\rangle _{n_{0}}=\frac{1}{\mathcal{H}}\sum_{k>0}\frac{1}{%
(e^{\varepsilon _{k}/T}-1)^{2}}.  \label{II38}
\end{equation}

At arbitrary temperatures, the condensate master equation (\ref{II20})
contains two non-trivial parameters, $\mathcal{H}$ and $\eta $,%
\[
\dot{p}_{n_{0}}=-\kappa \{(1+\eta
)[(N-n_{0})(n_{0}+1)p_{n_{0}}-(N-n_{0}+1)n_{0}p_{n_{0}-1}]+ 
\]%
\begin{equation}
\lbrack \mathcal{H}+(N-n_{0})\eta ]n_{0}p_{n_{0}}-[\mathcal{H}%
+(N-n_{0}-1)\eta ](n_{0}+1)p_{n_{0}+1}\}.  \label{II39}
\end{equation}

The steady-state solution of Eq. (\ref{II39}) is given by 
\begin{equation}
p_{n_{0}}=\frac{1}{Z_{N}}\frac{(N-n_{0}+\mathcal{H}/\eta -1)!}{(\mathcal{H}%
/\eta -1)!(N-n_{0})!}\Bigl(\frac{\eta }{1+\eta }\Bigr)^{N-n_{0}},
\label{II40}
\end{equation}%
where the canonical partition function $Z_{N}=1/p_{N}$ is 
\begin{equation}
Z_{N}=\sum_{n_{0}=0}^{N}\left( 
\begin{array}{c}
N-n_{0}+\mathcal{H}/\eta -1 \\ 
N-n_{0}%
\end{array}%
\right) \Bigl(\frac{\eta }{1+\eta }\Bigr)^{N-n_{0}}.  \label{II41}
\end{equation}

\begin{figure}[h]
\begin{center}
\includegraphics[width=9cm]{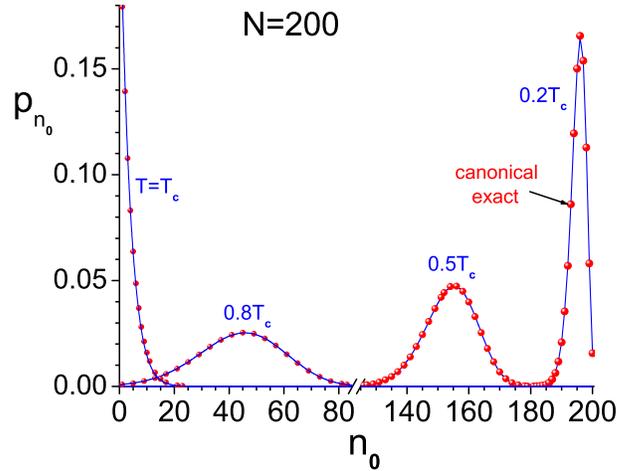}
\end{center}
\caption{Probability distribution of the ground-state occupation, $p_{n_0}$,
at the temperature $T=0.2$, $0.5$, $0.8$ and $1.0$ $T_c$ in an isotropic
harmonic trap with $N=200$ atoms as calculated from the solution of the
condensate master equation (\protect\ref{II20}) in the quasithermal
approximation, Eq. (\protect\ref{II40}), (solid line) and the ``exact"
numerical dots obtained in the canonical ensemble.}
\label{Npn0}
\end{figure}

The master equation (\ref{II39}) for $p_{n_{0}}$, and the analytic
approximate expressions (\ref{II40}) and (\ref{II41}) for the condensate
distribution function $p_{n_{0}}$ and the partition function $Z_{N}$,
respectively, are among the main results of the condensate master equation
approach. Now we are able to present the key picture of the theory of BEC
fluctuations, that is the probability distribution $p_{n_{0}}$, Fig. \ref%
{Npn0}. Analogy with the evolution of the photon number distribution in a
laser mode (from thermal to coherent, lasing) is obvious from a comparison
of Fig. \ref{Npn0} and Fig. \ref{las3}. With an increase of the number of
atoms in the trap, $N$, the picture of the ground-state occupation
distribution remains qualitatively the same, just a relative width of all
peaks becomes narrower.

\begin{figure}[h]
\begin{center}
\includegraphics[width=15cm]{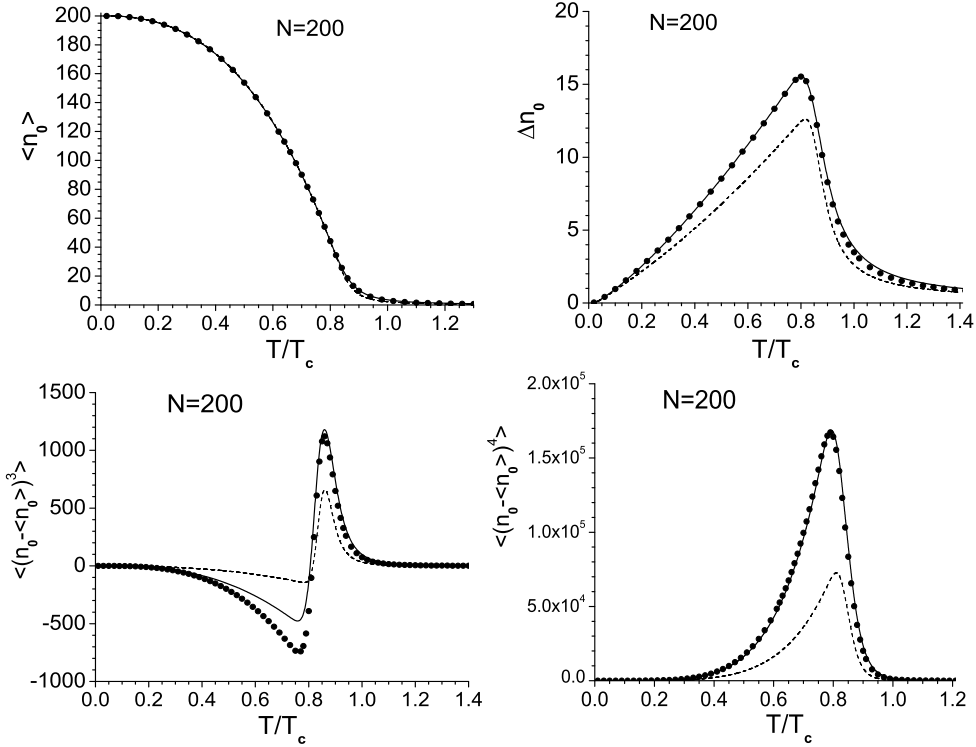}
\end{center}
\caption{The first four central moments for the ideal Bose gas in an
isotropic harmonic trap with $N=200$ atoms as calculated via the solution of
the condensate master equation (solid lines - quasithermal approximation,
Eq. (\protect\ref{II40}); dash lines - low temperature approximation, Eq. (%
\protect\ref{II30})). Dots are ``exact" numerical result obtained in the
canonical ensemble.}
\label{m1234}
\end{figure}

%\begin{figure}[tbp]
%\center \epsfxsize=15cm\epsffile{m1234.eps}
%\caption{The first four central moments for the ideal Bose gas in an
%isotropic harmonic trap with $N=200$ atoms as calculated via the solution of
%the condensate master equation (solid lines - quasithermal approximation,
%Eq. (\protect\ref{II40}); dash lines - low temperature approximation, Eq. (%
%\protect\ref{II30})). Dots are ``exact" numerical result obtained in the
%canonical ensemble.}
%\label{m1234}
%\end{figure}

%\begin{figure}[tbp]
%\center \epsfxsize=9cm\epsffile{pn0r.eps}
%\caption{Probability distribution of the ground-state occupation, $p_{n_0}$,
%at the temperature $T=0.2$, $0.5$, $0.8$ and $1.0$ $T_c$ in an isotropic
%harmonic trap with $N=200$ atoms as calculated from the solution of the
%condensate master equation (\protect\ref{II20}) in the quasithermal
%approximation, Eq. (\protect\ref{II40}), (solid line) and the ``exact"
%numerical dots obtained in the canonical ensemble.}
%\label{Npn0}
%\end{figure}

The average number of atoms condensed in the ground state of the trap is 
\begin{equation}
\langle n_{0}\rangle =N-\mathcal{H}+p_{0}\eta (N+\mathcal{H}/\eta )\quad .
\label{II51}
\end{equation}
The squared variance and higher central moments can be also calculated
analytically, e.g., 
\begin{equation}
\Delta n_{0}^{2}=(1+\eta )\mathcal{H}-p_{0}(\eta N+\mathcal{H})\Bigl(N-%
\mathcal{H}+1+\eta \Bigr)-p_{0}^{2}(\eta N+\mathcal{H})^{2},  \label{II54}
\end{equation}%
where 
\begin{equation}
p_{0}=\frac{1}{Z_{N}}\frac{(N+\mathcal{H}/\eta -1)!}{N!(\mathcal{H}/\eta -1)!%
}\left( \frac{\eta }{1+\eta }\right) ^{N}  \label{II55}
\end{equation}%
is the probability that there are no atoms in the condensate.

The first four central moments for the Bose gas in a harmonic trap with $%
N=200$ atoms are presented in Fig. \ref{m1234} as the functions of
temperature in different approximations. It is clearly seen that the
analytic results based on quasithermal distribution are indistinguishable
from the exact numerical results for the mean, and the second and the fourth
moments. The results for the third moment are however quantitatively
somewhat different. The success of the master equation approach is that the
analytic expressions are available for the partition function as well as the
condensate distribution function that mimic the exact solution to a
remarkable degree for a mesoscopic ideal Bose gas.

As a final point, we mention that a laser phase transition analogy exists
via the $P$-representation of the density matrix [DeGiorgio and Scully
(1970), Graham and Haken (1970)]

\begin{equation}
\rho =\int \frac{d^{2}\alpha }{\pi }P(\alpha ,\alpha ^{\ast })|\alpha
><\alpha |,  \label{P-rep}
\end{equation}%
where $|\alpha >$ is a coherent state. The steady-state solution of the
Fokker-Planck equation for laser near threshold is [Scully and Zubairy
(1997)] 
\begin{equation}
P(\alpha ,\alpha ^{\ast })=\frac{1}{\mathcal{N}}\exp \left[ \left( \frac{%
\mathcal{A}-\mathcal{C}}{\mathcal{A}}\right) |\alpha |^{2}-\frac{\mathcal{B}%
}{2\mathcal{A}}|\alpha |^{4}\right]  \label{P rep laser}
\end{equation}%
which clearly indicates a formal similarity between 
\begin{equation}
\ln P(\alpha ,\alpha ^{\ast })=-\ln \mathcal{N}+(1-\mathcal{H}%
/(N+1))n_{0}-(1/2(N+1))n_{0}^{2}  \label{lnP}
\end{equation}%
for the laser equation and the Ginzburg-Landau type free energy [Scully and
Zubairy (1997), DeGiorgio and Scully (1970), Graham and Haken (1970)] 
\begin{equation}
G(n_{0})=\ln p_{n_{0}}\approx const+a(T)n_{0}+b(T)n_{0}^{2},  \label{II29}
\end{equation}%
where $|\alpha |^{2}=n_{0},$ $a(T)=-(N-\mathcal{H})/N$ and $b(T)=1/2N$ for
large $N$ near $T_{c}$.

\subsection{Squeezing, Noise Reduction and BEC fluctuations}

The term \textquotedblleft squeezing" originates from the studies of a noise
reduction in quantum optics. In the present BEC context this aspect of
(quantum) optical physics is relevant to the characteristic function for the
total number of atoms in the two, $\mathbf{k}$ and $-\mathbf{k}$, modes
squeezed by Bogoliubov coupling. A detailed derivation of the characteristic
function for the fluctuations of the number of atoms in the two excited
modes squeezed by the Bogoliubov coupling is presented in [Kocharovsky et
al. (2000b)], it utilizes known results for the squeezed states of the
radiation field and is given by%
\[
\Theta _{\pm \mathbf{k}}(u)\equiv Tr\left\{ e^{iu(\hat{\beta}_{\mathbf{k}%
}^{+}\hat{\beta}_{\mathbf{k}}+\hat{\beta}_{-\mathbf{k}}^{+}\hat{\beta}_{-%
\mathbf{k}})}e^{-\varepsilon _{\mathbf{k}}(\hat{b}_{\mathbf{k}}^{+}\hat{b}_{%
\mathbf{k}}+\hat{b}_{-\mathbf{k}}^{+}\hat{b}_{-\mathbf{k}})/T}\left(
1-e^{-\varepsilon _{\mathbf{k}}/T}\right) ^{2}\right\} 
\]
\begin{equation}
=\frac{(z(A_{\mathbf{k}})-1)(z(-A_{\mathbf{k}})-1)}{(z(A_{\mathbf{k}%
})-e^{iu})(z(-A_{\mathbf{k}})-e^{iu})},  \label{III68}
\end{equation}%
where 
\begin{equation}
A_{\mathbf{k}}=\frac{V}{\bar{n}_{0}U_{\mathbf{k}}}\left( \varepsilon _{%
\mathbf{k}}-\frac{\hbar ^{2}\mathbf{k}^{2}}{2M}-\frac{\bar{n}_{0}U_{\mathbf{k%
}}}{V}\right) ,\qquad z(A_{\mathbf{k}})=\frac{A_{\mathbf{k}}-e^{\varepsilon
_{\mathbf{k}}/T}}{A_{\mathbf{k}}e^{\varepsilon _{\mathbf{k}}/T}-1},
\end{equation}%
$\varepsilon _{\mathbf{k}}$ is the energy of Bogoliubov quasiparticles%
\begin{equation}
\varepsilon _{\mathbf{k}}=\sqrt{\left( \frac{\hbar ^{2}\mathbf{k}^{2}}{2M}+%
\frac{\bar{n}_{0}U_{\mathbf{k}}}{V}\right) ^{2}-\left( \frac{\bar{n}_{0}U_{%
\mathbf{k}}}{V}\right) ^{2}},  \label{III65}
\end{equation}%
$M$ is the atomic mass, $V$ is the condensate volume, $U_{\mathbf{k}}$ is
the atom-atom scattering energy. In Eq. (\ref{III68}) $\hat{\beta}_{\mathbf{k%
}}$ are bare canonical ensemble quasiparticles which are related to
Bogoliubov quasiparticles $\hat{b}_{\mathbf{k}}$ by the canonical
transformation 
\begin{equation}
\hat{\beta}_{\mathbf{k}}=u_{\mathbf{k}}\hat{b}_{\mathbf{k}}+v_{\mathbf{k}}%
\hat{b}_{-\mathbf{k}}^{+},\quad \hat{\beta}_{\mathbf{k}}^{+}=u_{\mathbf{k}}%
\hat{b}_{\mathbf{k}}^{+}+v_{\mathbf{k}}\hat{b}_{-\mathbf{k}},
\end{equation}%
where $u_{k}$ and $v_{k}$ are Bogoliubov amplitudes%
\begin{equation}
u_{\mathbf{k}}=\frac{1}{\sqrt{1-A_{\mathbf{k}}^{2}}},\quad v_{\mathbf{k}}=%
\frac{A_{\mathbf{k}}}{\sqrt{1-A_{\mathbf{k}}^{2}}}.
\end{equation}

The characteristic function for the distribution of the total number of the
excited atoms is equal to the product of the coupled-mode characteristic
functions, $\Theta _{n}(u)\quad =\quad \Pi _{\mathbf{k}\neq 0,mod\{\pm 
\mathbf{k}\}}\Theta _{\pm \mathbf{k}}(u)$, since different pairs of $(%
\mathbf{k},-\mathbf{k})$-modes are independent to the first approximation.
The product $\Pi $ runs over all different pairs of $(\mathbf{k},-\mathbf{k}%
) $-modes.

By doing all calculations via the canonical-ensemble quasiparticles we
automatically take into account all correlations introduced by the
canonical-ensemble constraint. The important conclusion is that for square
well trap the ground state occupation fluctuations are not Gaussian even in
the thermodynamic limit. It is more convenient, in particular, for the
analysis of the non-Gaussian properties, to solve for the cumulants $\kappa
_{m}$ which are defined as coefficients in Taylor expansion $\ln \Theta
_{n}(u)=\sum_{m=1}^{\infty }\kappa _{m}(iu)^{m}/m!$, where $\Theta _{n}(u)$
is the characteristic function $\Theta _{n}(u)=Tr\left\{ e^{iu\hat{n}}\hat{%
\rho}\right\} $. There are simple relations between $\kappa _{m}$ and
central moments $\mu _{m}$, in particular, 
\begin{equation}
\kappa _{1}=\bar{n},\quad \kappa _{2}=\mu _{2},\quad \kappa _{3}=\mu
_{3},\quad \kappa _{4}=\mu _{4}-3\mu _{2}^{2}.
\end{equation}%
The \textquotedblleft generating cumulants" $\tilde{\kappa}_{m}$ are simply
related to the cumulants $\kappa _{m}$ by 
\begin{equation}
\kappa _{1}=\tilde{\kappa}_{1},\quad \kappa _{2}=\tilde{\kappa}_{2}+\tilde{%
\kappa}_{1},\quad \kappa _{3}=\tilde{\kappa}_{3}+3\tilde{\kappa}_{2}+\tilde{%
\kappa}_{1},\quad \kappa _{4}=\tilde{\kappa}_{4}+6\tilde{\kappa}_{3}+7\tilde{%
\kappa}_{2}+\tilde{\kappa}_{1}.
\end{equation}%
For Gaussian distribution $\kappa _{m}=0$, for $m=3,4,\ldots $.

The explicit formula for all cumulants in the dilute weakly interacting Bose
gas was obtained in [Kocharovsky et al. (2000b)]:%
\begin{equation}
\tilde{\kappa}_{m}=\frac{1}{2}(m-1)!\sum_{\mathbf{k}\neq 0}\left[ \frac{1}{%
\left( z(A_{\mathbf{k}})-1\right) ^{m}}+\frac{1}{\left( z(-A_{\mathbf{k}%
})-1\right) ^{m}}\right] .  \label{III71}
\end{equation}%
For the ideal gas the answer is%
\begin{equation}
\tilde{\kappa}_{m}=(m-1)!\sum_{\mathbf{k}\neq 0}\frac{1}{(e^{\varepsilon _{%
\mathbf{k}}/T}-1)^{m}}.  \label{idg}
\end{equation}

In comparison with the ideal Bose gas, Eq. (\ref{idg}), for the interacting
particles we have effectively a mixture of two species of atom pairs with $%
z(\pm A_{\mathbf{k}})$ instead of $\exp (\varepsilon _{\mathbf{k}}/T)$.

\begin{figure}[h]
\begin{center}
\includegraphics[width=5.5cm]{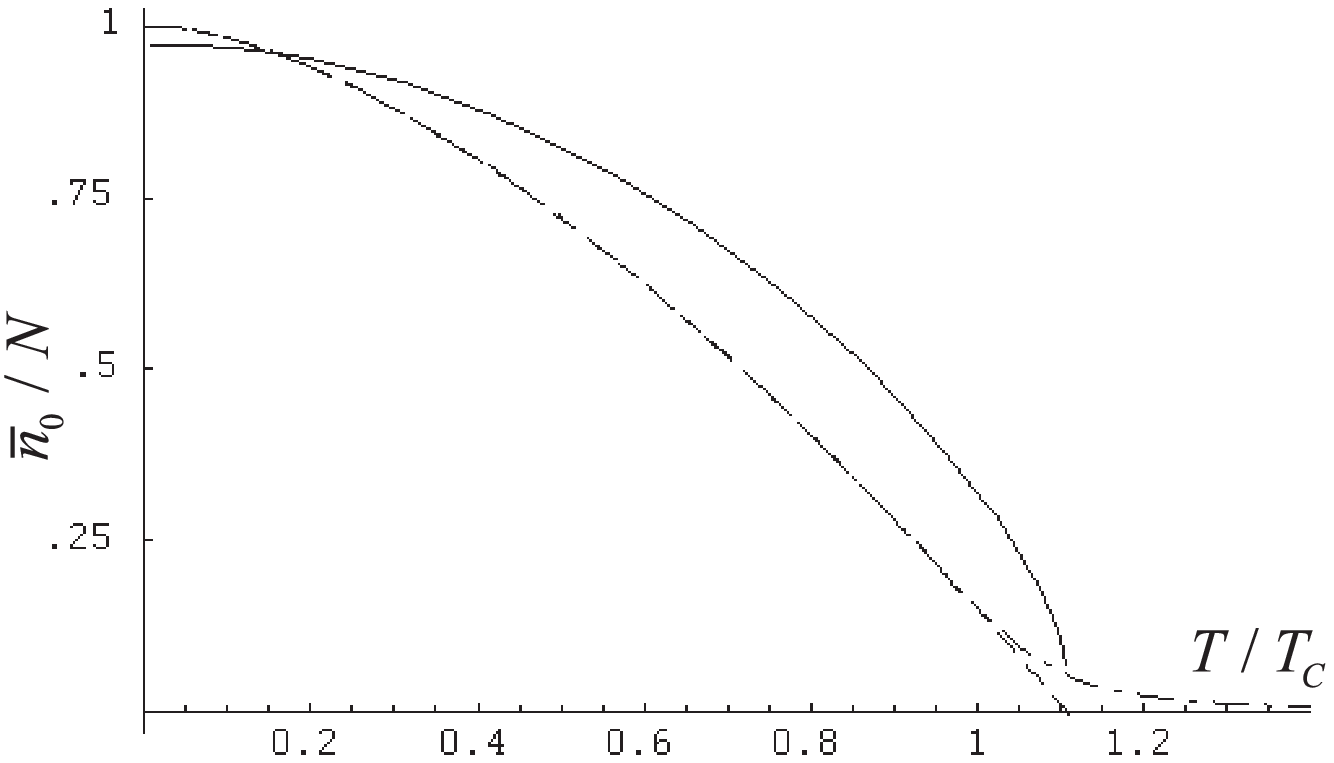} %
\includegraphics[width=5.5cm]{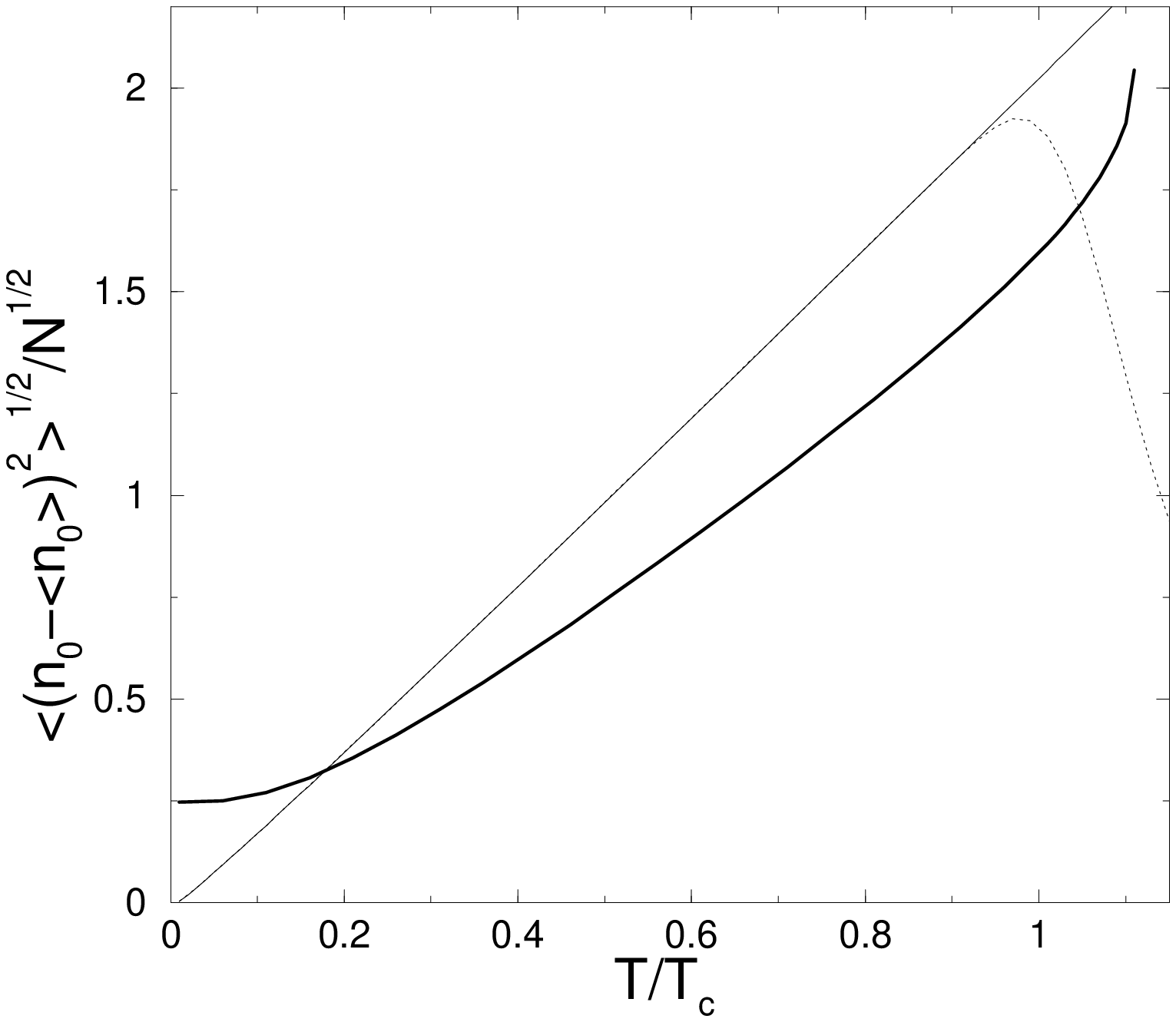} %
\includegraphics[width=5.5cm]{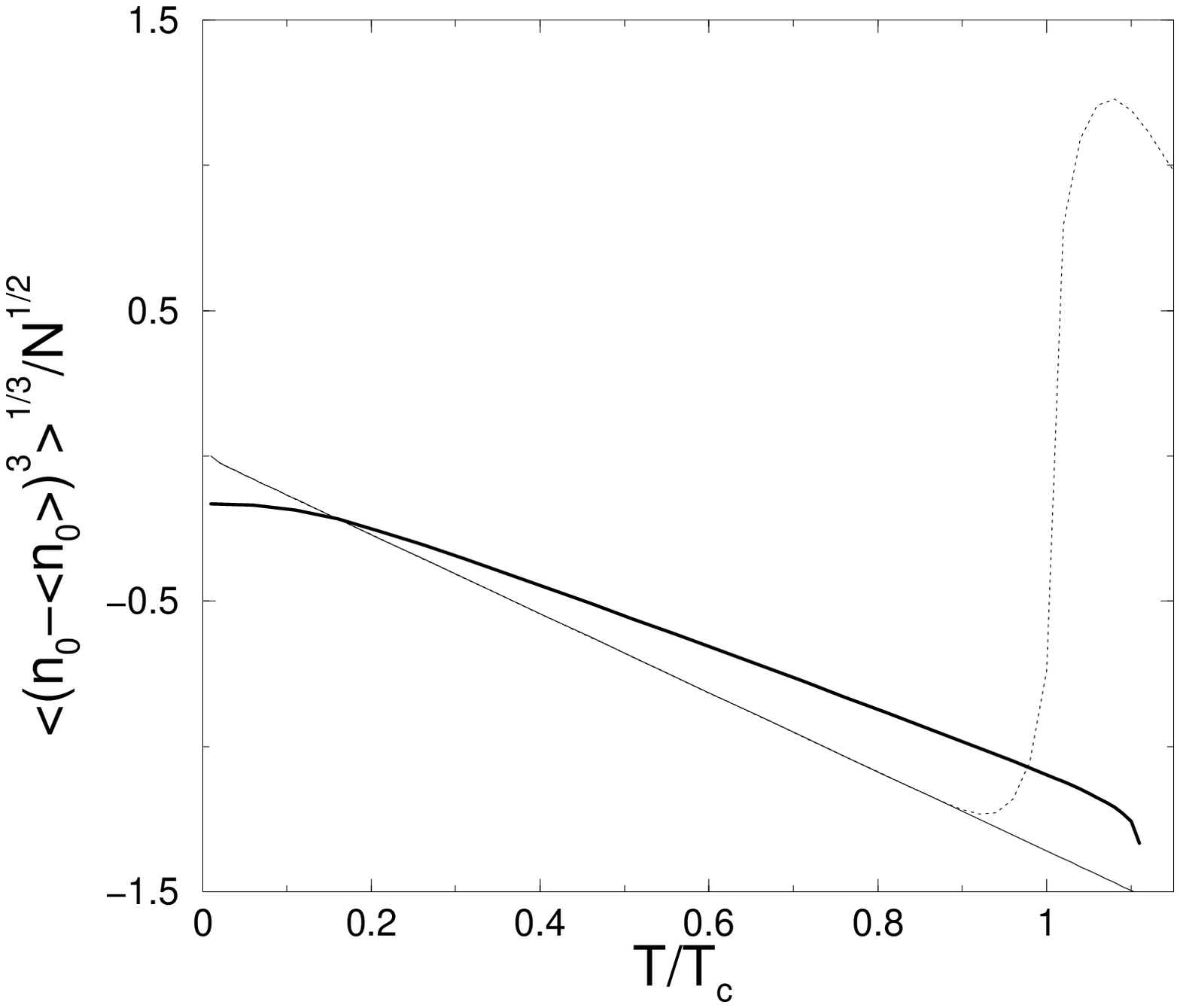} %
\includegraphics[width=5.5cm]{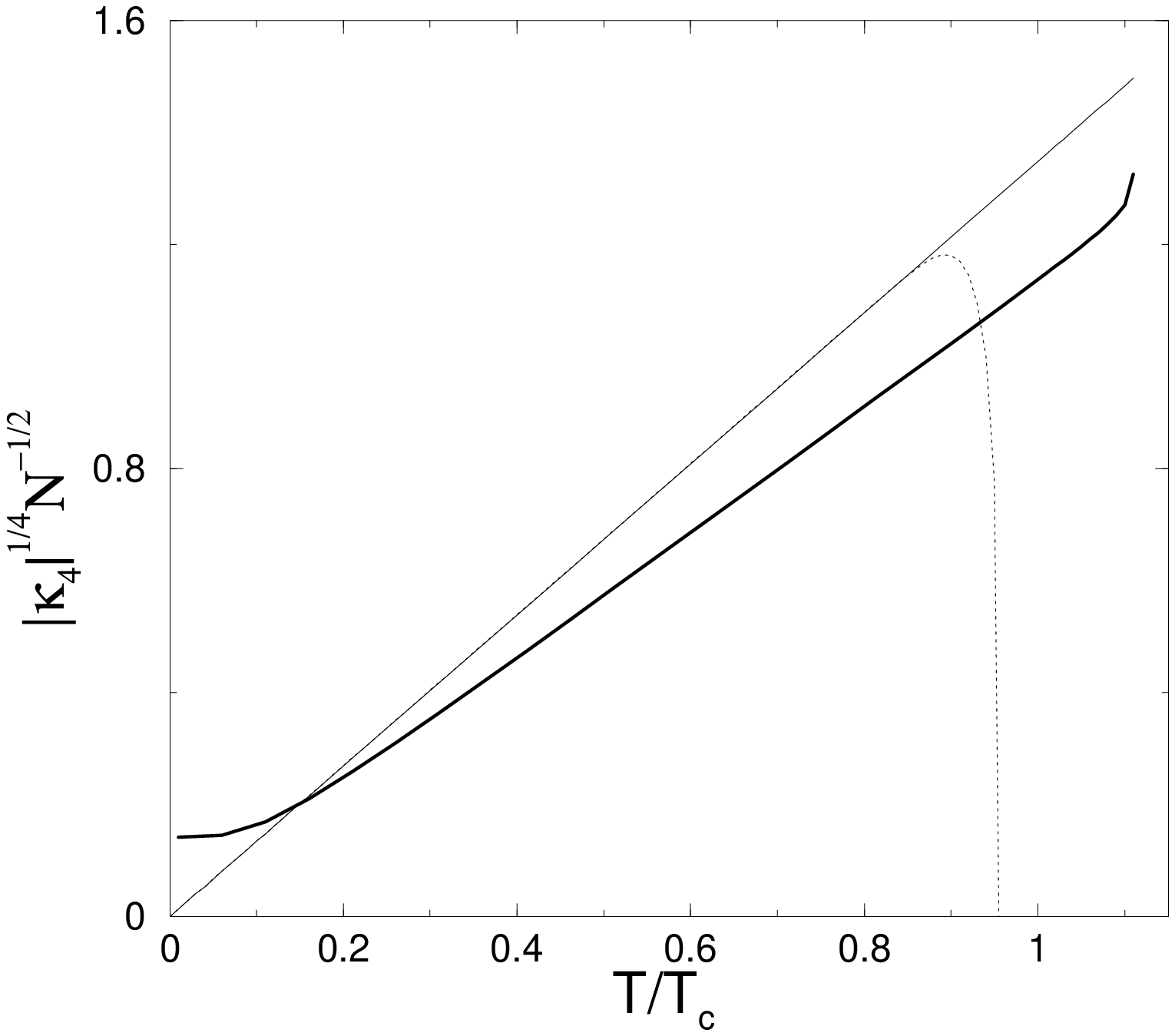}
\end{center}
\caption{Temperature scaling of the first four cumulants, the mean value $%
\bar{n}_0/N = N-\protect\kappa_1/N$, the variance $\protect\sqrt{\protect%
\kappa_2/N} = \langle(n_0-\bar{n}_0)^2\rangle^{1/2}/N^{1/2}$, the third
central moment $-\protect\kappa_3^{1/3}/N^{1/2} = \langle(n_0-\bar{n}%
_0)^3\rangle^{1/3}/N^{1/2}$, the fourth cumulant $|\protect\kappa%
_4|^{1/4}/N^{1/2} = |\langle(n_0-\bar{n}_0)^4\rangle-3\protect\kappa%
_2^2|/N^2 $, of the ground-state occupation fluctuations for the dilute
weakly interacting Bose gas with $U_0 N^{1/3} /\protect\varepsilon_1 V =
0.05 $ (thick solid lines), as compared with the ideal gas (thin solid
lines) and with the ``exact" numerical result in the canonical ensemble
(dot-dashed lines) for the ideal gas in the box; $N=1000 $. For the ideal
gas the thin solid lines are almost indistinguishable from the ``exact"
dot-dashed lines in the condensed region, $T<T_c(N)$. Temperature is
normalized by the standard thermodynamic-limit critical value $T_c$ $(
N=\infty )$ that differs from the finite-size value $T_c (N)$, as is clearly
seen in graphs.}
\label{FigIII2}
\end{figure}

It is important to emphasize that the first equation in (\ref{III71}), $m=1$%
, is a non-linear self-consistency equation, 
\begin{equation}
N-\bar{n}_{0}=\kappa _{1}(\bar{n}_{0})\equiv \sum_{\mathbf{k}\neq 0}\frac{%
1+A_{\mathbf{k}}^{2}e^{\varepsilon _{\mathbf{k}}/T}}{(1-A_{\mathbf{k}%
}^{2})(e^{\varepsilon _{\mathbf{k}}/T}-1)},  \label{III72}
\end{equation}%
to be solved for the mean number of ground-state atoms $\bar{n}_{0}(T)$,
since the Bogoliubov coupling coefficient $A_{\mathbf{k}}$, and the energy
spectrum $\varepsilon _{\mathbf{k}}$, are themselves functions of the mean
value $\bar{n}_{0}$. Then, all the other equations in (\ref{III71}), $m\geq
2 $, are nothing else but explicit expressions for all cumulants, $m\geq 2$,
if one substitutes the solution of the self-consistency equation (\ref{III72}%
) for the mean value $\bar{n}_{0}$. The Eq. (\ref{III72}), obtained for the
interacting Bose gas in the canonical-ensemble quasiparticle approach,
coincides precisely with the self-consistency equation for the
grand-canonical dilute gas in the so-called first-order Popov approximation
(see a review in [Shi and Griffin (1998)]). The latter is well established
as a reasonable first approximation for the analysis of the
finite-temperature properties of the dilute Bose gas and is not valid only
in a very small interval near $T_{c}$, given by $T_{c}-T<a(N/V)^{1/3}T_{c}%
\ll T_{c}$, where $a=MU_{0}/4\pi \hbar ^{2}$ is a usual $s$-wave scattering
length. The analysis of the Eq.~(\ref{III72}) shows that in the dilute gas
the self-consistent value $\bar{n}_{0}(T)$ is close to that given by the
ideal gas model, and for very low temperatures goes smoothly to the value
given by the standard Bogoliubov theory [Lifshitz and Pitaevskii (1981),
Abrikosov et al. (1963), Fetter and Walecka (1971)] for a small condensate
depletion, $N-\bar{n}_{0}\ll N$. This is illustrated by Fig. \ref{FigIII2}
in which we show the first four cumulants. Near the critical temperature $%
T_{c}$ the number of excited quasiparticles is relatively large, so that
along with the Bogoliubov coupling other, higher order effects of
interaction should be taken into account to get a complete theory. Note that
the effect of a weak interaction on the condensate fluctuations is very
significant, see Fig. \ref{FigIII2}, even if the mean number of condensed
atoms changes by relatively small amount.

%\begin{figure}[tbp]
%\center\epsfxsize=5.5cm\epsffile{Pic1.eps} \center\epsfxsize=5.5cm%
%\epsffile{Cnb3m2.eps} \center\epsfxsize=5.5cm\epsffile{Cnb3m3.eps} \center%
%\epsfxsize=5.5cm\epsffile{M4n.eps}
%\caption{Temperature scaling of the first four cumulants, the mean value $%
%\bar{n}_0/N = N-\protect\kappa_1/N$, the variance $\protect\sqrt{\protect%
%\kappa_2/N} = \langle(n_0-\bar{n}_0)^2\rangle^{1/2}/N^{1/2}$, the third
%central moment $-\protect\kappa_3^{1/3}/N^{1/2} = \langle(n_0-\bar{n}%
%_0)^3\rangle^{1/3}/N^{1/2}$, the fourth cumulant $|\protect\kappa%
%_4|^{1/4}/N^{1/2} = |\langle(n_0-\bar{n}_0)^4\rangle-3\protect\kappa%
%_2^2|/N^2 $, of the ground-state occupation fluctuations for the dilute
%weakly interacting Bose gas with $U_0 N^{1/3} /\protect\varepsilon_1 V =
%0.05 $ (thick solid lines), as compared with the ideal gas (thin solid
%lines) and with the ``exact" numerical result in the canonical ensemble
%(dot-dashed lines) for the ideal gas in the box; $N=1000 $. For the ideal
%gas the thin solid lines are almost indistinguishable from the ``exact"
%dot-dashed lines in the condensed region, $T<T_c(N)$. Temperature is
%normalized by the standard thermodynamic-limit critical value $T_c (
%N=\infty )$ that differs from the finite-size value $T_c (N)$, as is clearly
%seen in graphs.}
%\label{FigIII2}
%\end{figure}

\section{Hybrid approach to condensate fluctuations}

\setcounter{equation}{0}

We now show how to combine ideas from the canonical ensemble quasiparticle
formalism of [Kocharovsky et al. (2000b)] (which works well for an
interacting gas at temperature not too close to $T_{c}$ when $\sqrt{\mu _{2}}%
\ll \bar{n}_{0}$) with the physics of the master equation approach (in the
spirit of the quantum theory of the laser) [Kocharovsky et al. (2000a)], in
order to obtain essentially perfect quantitative agreement with the exact
numerical solution of the canonical partition function at \textit{all}
temperatures for the fluctuation statistics of the Bose gas. Such a hybrid
technique was proposed in [Svidzinsky and Scully (2006)].

We recall the master equation for the condensate probability distribution
for a non-interacting Bose gas ((\ref{II20}) 
\begin{equation}
\frac{1}{\kappa }\dot{p}%
_{n_{0}}=-K_{n_{0}}(n_{0}+1)p_{n_{0}}+K_{n_{0}-1}n_{0}p_{n_{0}-1}-
H_{n_{0}}n_{0}p_{n_{0}}+H_{n_{0}+1}(n_{0}+1)p_{n_{0}+1},
\end{equation}%
The detailed balance condition yields 
\begin{equation}
\frac{p_{n_{0}+1}}{p_{n_{0}}}=\frac{K_{n_{0}}}{H_{n_{0}+1}}.  \label{p1}
\end{equation}

Since the occupation number of the ground state cannot be larger than $N$
there is a canonical ensemble constraint $p_{N+1}=0$ and, hence, $K_{N}=0$.
In contrast to $p_{n_{0}}$, the ratio $p_{n_{0}+1}/p_{n_{0}}$ as a function
of $n_{0}$ shows simple monotonic behavior. We approximate $K_{n_{0}}$ and $%
H_{n_{0}}$ by a few terms of the Taylor expansion near the point $n_{0}=N$ 
\begin{equation}
K_{n_{0}}=(N-n_{0})(1+\eta )+\alpha (N-n_{0})^{2},  \label{p2}
\end{equation}%
\begin{equation}
H_{n_{0}}=\mathcal{H}+(N-n_{0})\eta +\alpha (N-n_{0})^{2}.  \label{p3}
\end{equation}%
The parameters $\mathcal{H}$, $\eta $ and $\alpha $ are independent of $%
n_{0} $; they are functions of the occupation of the excited levels. We
derive them below by matching the first three central moments in the low
temperature limit with the result of [Kocharovsky et al. (2000b)]. We note
that the detailed balance equation (\ref{p1}) is the Pad\'{e} approximation
[Baker (1996)] of the function $p_{n_{0}+1}/p_{n_{0}}$. Pad\'{e} summation
has proven to be useful in many applications, including condensed-matter
problems and quantum field theory.

Equations (\ref{p1})-(\ref{p3}) yield an analytical expression for the
condensate distribution function 
\begin{equation}
p_{n_{0}}=\frac{1}{{\mathcal{Z}_{N}}}\frac{%
(N-n_{0}-1+x_{1})!(N-n_{0}-1+x_{2})!}{(N-n_{0})!(N-n_{0}+(1+\eta )/\alpha )!}%
,  \label{p4}
\end{equation}%
where $x_{1,2}=(\eta \pm \sqrt{\eta ^{2}-4\alpha \mathcal{H}})/2\alpha $ and 
$\mathcal{Z}_{N}$ is the normalization constant determined by $%
\sum\limits_{n_{0}=0}^{N}$ $p_{n_{0}}=1$. In the particular case $\eta
=\alpha =0$ Eq. (\ref{p4}) reduces to Eq. (\ref{II30}) obtained in the low
temperature approximation.

Using the distribution function (\ref{p4}) we find that, in the validity
range of [Kocharovsky et al. (2000b)] (at low enough $T$), the first three
central moments $\mu _{m}\equiv <(n_{0}-\bar{n}_{0})^{m}>$ are 
\begin{equation}
\bar{n}_{0}=N-\mathcal{H},\qquad \mu _{2}=(1+\eta )\mathcal{H}+\alpha 
\mathcal{H}^{2},  \label{p6}
\end{equation}%
\begin{equation}
\mu _{3}=-\mathcal{H}(1+\eta +\alpha \mathcal{H})(1+2\eta +4\alpha \mathcal{H%
}).  \label{p8}
\end{equation}%
Eqs. (\ref{p6}), (\ref{p8}) thus yield%
\begin{equation}
\mathcal{H}=N-\bar{n}_{0},\qquad \eta =\frac{1}{2}\left( \frac{\mu _{3}}{\mu
_{2}}-3+\frac{4\mu _{2}}{\mathcal{H}}\right) ,  \label{p9}
\end{equation}%
\begin{equation}
\alpha =\frac{1}{\mathcal{H}}\left( \frac{1}{2}-\frac{\mu _{2}}{\mathcal{H}}-%
\frac{\mu _{3}}{2\mu _{2}}\right) .  \label{p11}
\end{equation}

On the other hand, the result of [Kocharovsky et al. (2000b)] for an
interacting Bogoliubov gas is (see Appendix for derivation of $\bar{n}_{0}$
and $\mu _{2}$) 
\begin{equation}
\bar{n}_{0}=N-\sum\limits_{k\neq 0}\left[ \left( u_{k}^{2}+v_{k}^{2}\right)
f_{k}+v_{k}^{2}\right] ,  \label{a1}
\end{equation}%
\begin{equation}
\mu _{2}=\sum\limits_{k\neq 0}\left[
(1+8u_{k}^{2}v_{k}^{2})(f_{k}^{2}+f_{k})+2u_{k}^{2}v_{k}^{2}\right] ,
\label{a2}
\end{equation}%
\begin{equation}
\mu _{3}=-\sum\limits_{k\neq 0}\left( u_{k}^{2}+v_{k}^{2}\right) \left[
(1+16u_{k}^{2}v_{k}^{2})(2f_{k}^{3}+3f_{k}^{2}+f_{k})+\right. \left.
4u_{k}^{2}v_{k}^{2}(1+2f_{k})\right] ,  \label{a3}
\end{equation}%
where $f_{k}=1/[\exp (\varepsilon _{k}/k_{B}T)-1]$ is the number of
elementary excitations with energy $\varepsilon _{k}$ present in the system
at thermal equilibrium, $u_{k}$ and $v_{k}$ are Bogoliubov amplitudes.
Substitute for $\bar{n}_{0}$, $\mu _{2}$ and $\mu _{3}$ in Eqs. (\ref{p9}), (%
\ref{p11}) their expressions [Kocharovsky et al. (2000b)] (\ref{a1})-(\ref%
{a3}) yields the unknown parameters $\mathcal{H}$, $\eta $ and $\alpha $.
The beauty of the present \textquotedblleft matched asymptote" derivation is
that the formulas for $\mathcal{H}$, $\eta $ and $\alpha $\ are applicable
at all temperatures, i.e. not only in the validity range of [Kocharovsky et
al. (2000b)]. The distribution function (\ref{p4}) together with Eqs. (\ref%
{p9}), (\ref{p11}) provides complete knowledge of the condensate statistics
at all $T$. Taking $v_{k}=0$ and $u_{k}=1$ in (\ref{a1})-(\ref{a3}) we
obtain the ideal gas limit.

\begin{figure}[h]
\begin{center}
\includegraphics[width=13cm]{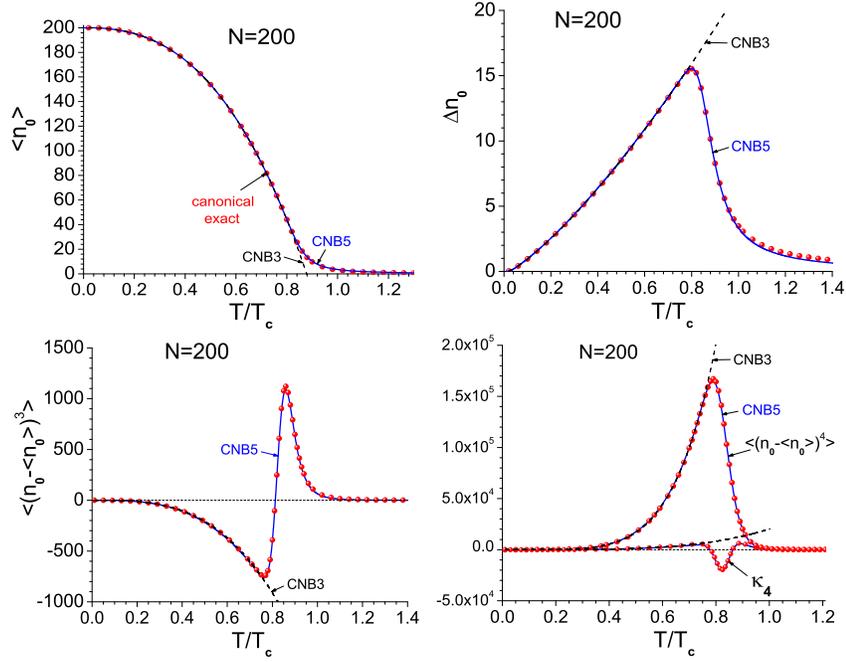}
\end{center}
\caption{ Average condensate particle number $<n_0>$, its variance $\Delta
n_0=\protect\sqrt{<(n_0-\bar n_0)^2>}$, third and fourth central moments $%
<(n_{0}-\bar{n}_{0})^{m}>$ ($m=3,4$) and fourth cumulant $\protect\kappa_4$
as a function of temperature for an ideal gas of $N=200$ particles in a
harmonic trap. Solid lines (CNB5) show the result of the hybrid approach.
[Kocharovsky et al. (2000b)] yields dashed lines (CNB3). Dots are ``exact"
numerical simulation in the canonical ensemble. The temperature is
normalized by the thermodynamic critical temperature for the trap $%
T_{c}=\hbar \protect\omega N^{1/3}/k_{B}\protect\zeta (3)^{1/3}$, where $%
\protect\omega $ is the trap frequency. }
\label{n14trap}
\end{figure}

Figure \ref{n14trap} shows the average condensate particle number $\bar{n}%
_{0}$, its variance, third and fourth central moments $\mu _{m}$ and fourth
cumulant $\kappa _{4}$ as a function of $T$ for an ideal gas of $N=200$
particles in a harmonic trap. Solid lines are the result of the present
approach which is in remarkable agreement with the \textquotedblleft exact"
dots at all temperatures both for $\mu _{m}$ and $\kappa _{4}$. Central
moments and cumulants higher than fourth order are not shown here, but they
are also remarkably accurate at all temperatures. Results of [Kocharovsky et
al. (2000b)] are given by dashed lines which are accurate only at
sufficiently low $T$. Deviation of higher order cumulants ($m=3,4,\ldots $)
from zero indicates that the fluctuations are not Gaussian.

Clearly the present hybrid method passes the ideal gas test with flying
colors. We note the excellent agreement with the exact analysis for the
third central moment and fourth cumulant $\kappa _{4}$ given in Fig. \ref%
{n14trap}.

\begin{figure}[h]
\begin{center}
\includegraphics[width=13cm]{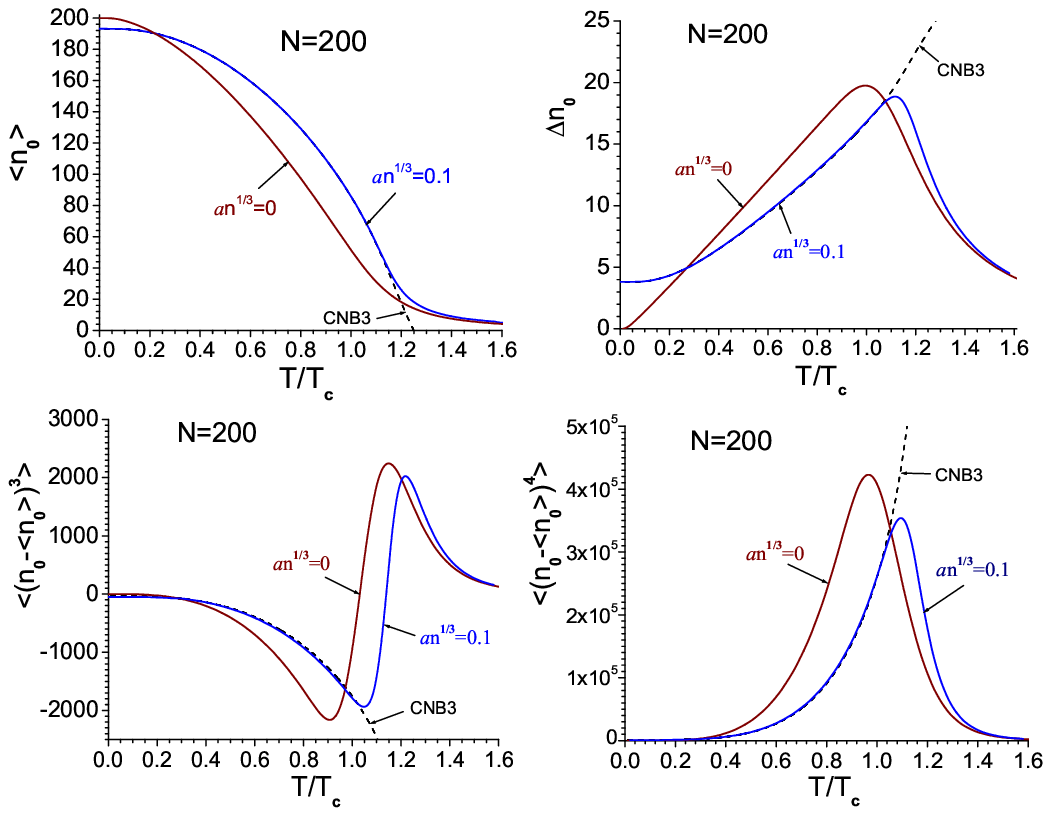}
\end{center}
\caption{ Average condensate particle number, its variance, third and fourth
central moments as a function of temperature for an ideal ($an^{1/3}=0$) and
interacting ($an^{1/3}=0.1$) Bose gas of $N=200$ particles in a box. Solid
lines are the result of the hybrid approach. [Kocharovsky et al. (2000b)]
yields dashed lines (CNB3). The temperature is normalized by the
thermodynamic critical temperature for the box $T_{c}=2\protect\pi \hbar^2
n^{2/3}/k_{B}M\protect\zeta (3/2)^{2/3}$, where $M$ is the particle mass. }
\label{n14box}
\end{figure}

Next we apply this technique to $N$ interacting Bogoliubov particles
confined in a box of volume $V$. The interactions are characterized by the
gas parameter $an^{1/3}$, where $a$ is the s-wave scattering length and $%
n=N/V$ is the particle density. The energy of Bogoliubov quasiparticles $%
\varepsilon _{k}$ depends on $\bar{n}_{0}$, hence, the equation $\bar{n}%
_{0}=\sum\limits_{n_{0}=0}^{N}$ $n_{0}p_{n_{0}}$ for $\bar{n}_{0}$ must be
solved self-consistently. In Fig. \ref{n14box} we plot $\bar{n}_{0}$, the
variance $\Delta n_{0}$, third and fourth central moments as a function of $%
T $ for an ideal and interacting ($an^{1/3}=0.1$) gas in the box. Solid
lines show the result of the present approach, while [Kocharovsky et al.
(2000b)] is represented by dashed lines. The present results agree well for
all $\mu _{m} $ with [Kocharovsky et al. (2000b)] in the range of its
validity. Near and above $T_{c} $ [Kocharovsky et al. (2000b)] becomes
inaccurate. However, the results of the present method are expected to be
accurate at all $T$. Indeed, in the limit $T\gg T_{c}$ the present results
(unlike [Kocharovsky et al. (2000b)]) merge with those for the ideal gas.
This is physically appealing since at high $T$ the kinetic energy becomes
much larger than the interaction energy and the gas behaves ideally. Similar
to the ideal gas, the interacting mesoscopic BEC $\bar{n}_{0}(T)$ exhibits a
smooth transition when passing through $T_{c}$.

One can see from Fig. \ref{n14box} that the repulsive interaction stimulates
BEC, and yields an increase in $\bar{n}_{0}$ at intermediate temperatures,
as compared to the ideal gas. This effect is known as \textquotedblleft
attraction in momentum space" and occurs for energetic reasons [Leggett
(2001)]. Bosons in different states interact more strongly than when they
are in the same state, and this favors multiple occupation of a single
one-particle state.

We gratefully acknowledge the support of the Office of Naval Research (Award
No. N00014-03-1-0385) and the Robert A. Welch Foundation (Grant No. A-1261).

%%%%%%%%%%%%%%%%%%%% Fig 3 %%%%%%%%%%%%%%%%%%%%%%%%%%%%%
%\begin{figure}[h]
%\bigskip
%%\centerline{\epsfxsize=1.1\textwidth\epsfysize=0.9\textwidth
%\center \epsfxsize=5.0cm\epsffile{n14trap.eps}
%%%%%%%%%%%%%%
%\epsfbox{n14trap.eps}}
%\caption{ Average condensate particle number $<n_0>$, its variance $\Delta
%n_0=\protect\sqrt{<(n_0-\bar n_0)^2>}$, third and fourth central moments $%
%<(n_{0}-\bar{n}_{0})^{m}>$ ($m=3,4$) and fourth cumulant $\protect\kappa_4$
%as a function of temperature for an ideal gas of $N=200$ particles in a
%harmonic trap. Solid lines show the result of the hybrid approach.
%[Kocharovsky et al. (2000b)] yields dashed lines. Dots are ``exact"
%numerical simulation in the canonical ensemble. The temperature is
%normalized by the thermodynamic critical temperature for the trap $%
%T_{c}=\hbar \protect\omega N^{1/3}/k_{B}\protect\zeta (3)^{1/3}$, where $%
%\protect\omega $ is the trap frequency. }
%\label{n14trap}
%\end{figure}
%%%%%%%%%%%%%%%%%%%%%%%%%%%%%%%%%%%%%%%%%%%%%%%%%%%%%%%%%

%%%%%%%%%%%%%%%%%%%% Fig  %%%%%%%%%%%%%%%%%%%%%%%%%%%%%
%\begin{figure}[h]
%\bigskip
%\centerline{\epsfxsize=1.1\textwidth\epsfysize=0.9\textwidth
%\epsfbox{n14box.eps}}
%\caption{ Average condensate particle number, its variance, third and fourth
%central moments as a function of temperature for an ideal ($an^{1/3}=0$) and
%interacting ($an^{1/3}=0.1$) Bose gas of $N=200$ particles in a box. Solid
%lines are the result of the hybrid approach. [Kocharovsky et al. (2000b)] yields
%dashed lines. The temperature is normalized by the thermodynamic
%critical temperature for the box $T_{c}=2\protect\pi \hbar^2 n^{2/3}/k_{B}M%
%\protect\zeta (3/2)^{2/3}$, where $M$ is the particle mass. }
%\label{n14box}
%\end{figure}
%%%%%%%%%%%%%%%%%%%%%%%%%%%%%%%%%%%%%%%%%%%%%%%%%%%%%%%%%

%\appendix

\begin{center}
{\large \textbf{Appendix}}
\end{center}

\section{Mean condensate particle number and its variance for weakly
interacting BEC}

\setcounter{equation}{0}

In the framework of Bogoliubov theory the particle operator can be expressed
in terms of the quasiparticle creation and annihilation operators as%
\begin{equation}
\hat{\beta}_{\mathbf{k}}=u_{\mathbf{k}}\hat{b}_{\mathbf{k}}+v_{\mathbf{k}}%
\hat{b}_{-\mathbf{k}}^{+},\quad \hat{\beta}_{\mathbf{k}}^{+}=u_{\mathbf{k}}%
\hat{b}_{\mathbf{k}}^{+}+v_{\mathbf{k}}\hat{b}_{-\mathbf{k}},
\end{equation}%
where $u_{k}$ and $v_{k}$ are Bogoliubov amplitudes. The total number of
particles out of the condensate is given by the expectation value of the
operator 
\begin{equation}
\hat{N}_{out}=\sum\limits_{k\neq 0}\hat{\beta}_{\mathbf{k}}^{+}\hat{\beta}_{%
\mathbf{k}}=\sum\limits_{k\neq 0}\left[ u_{\mathbf{k}}^{2}\hat{b}_{\mathbf{k}%
}^{+}\hat{b}_{\mathbf{k}}+v_{\mathbf{k}}^{2}\hat{b}_{\mathbf{k}}\hat{b}_{%
\mathbf{k}}^{+}+u_{\mathbf{k}}v_{\mathbf{k}}(\hat{b}_{\mathbf{k}}\hat{b}_{-%
\mathbf{k}}+\hat{b}_{\mathbf{k}}^{+}\hat{b}_{-\mathbf{k}}^{+})\right] .
\label{b1}
\end{equation}%
Using the particle number constraint $\hat{n}_{0}+\hat{N}_{out}=N$ we obtain 
\begin{equation}
\bar{n}_{0}=N-<\hat{N}_{out}>=N-\sum\limits_{k\neq 0}\left[ \left(
u_{k}^{2}+v_{k}^{2}\right) f_{k}+v_{k}^{2}\right] ,  \label{b2}
\end{equation}%
where $f_{k}=<\hat{b}_{\mathbf{k}}^{+}\hat{b}_{\mathbf{k}}>=1/[\exp
(\varepsilon _{k}/k_{B}T)-1]$ is the number of elementary excitations with
energy $\varepsilon _{k}$ present in the system at thermal equilibrium.

In a similar way one can calculate particle fluctuations [Giorgini et al.
(1998)]. Using Eq. (\ref{b1}) we have%
\[
\hat{N}_{out}^{2}=\sum\limits_{k\neq 0}\hat{\beta}_{\mathbf{k}}^{+}\hat{\beta%
}_{\mathbf{k}}\sum\limits_{q\neq 0}\hat{\beta}_{\mathbf{q}}^{+}\hat{\beta}_{%
\mathbf{q}}=\sum\limits_{k,q\neq 0}\left[ u_{\mathbf{k}}^{2}u_{\mathbf{q}%
}^{2}\hat{b}_{\mathbf{k}}^{+}\hat{b}_{\mathbf{k}}\hat{b}_{\mathbf{q}}^{+}%
\hat{b}_{\mathbf{q}}+v_{\mathbf{k}}^{2}v_{\mathbf{q}}^{2}\hat{b}_{\mathbf{k}}%
\hat{b}_{\mathbf{k}}^{+}\hat{b}_{\mathbf{q}}\hat{b}_{\mathbf{q}}^{+}+\right. 
\]%
\[
u_{\mathbf{k}}^{2}v_{\mathbf{q}}^{2}(\hat{b}_{\mathbf{k}}^{+}\hat{b}_{%
\mathbf{k}}\hat{b}_{\mathbf{q}}\hat{b}_{\mathbf{q}}^{+}+\hat{b}_{\mathbf{q}}%
\hat{b}_{\mathbf{q}}^{+}\hat{b}_{\mathbf{k}}^{+}\hat{b}_{\mathbf{k}})+u_{%
\mathbf{k}}^{2}u_{\mathbf{q}}v_{\mathbf{q}}\hat{b}_{\mathbf{k}}^{+}\hat{b}_{%
\mathbf{k}}(\hat{b}_{\mathbf{q}}\hat{b}_{-\mathbf{q}}+\hat{b}_{\mathbf{q}%
}^{+}\hat{b}_{-\mathbf{q}}^{+})+ 
\]%
\[
v_{\mathbf{k}}^{2}u_{\mathbf{q}}v_{\mathbf{q}}\hat{b}_{\mathbf{k}}\hat{b}_{%
\mathbf{k}}^{+}(\hat{b}_{\mathbf{q}}\hat{b}_{-\mathbf{q}}+\hat{b}_{\mathbf{q}%
}^{+}\hat{b}_{-\mathbf{q}}^{+})+u_{\mathbf{k}}v_{\mathbf{k}}u_{\mathbf{q}%
}^{2}(\hat{b}_{\mathbf{k}}\hat{b}_{-\mathbf{k}}+\hat{b}_{\mathbf{k}}^{+}\hat{%
b}_{-\mathbf{k}}^{+})\hat{b}_{\mathbf{q}}^{+}\hat{b}_{\mathbf{q}}+ 
\]%
\begin{equation}
\left. u_{\mathbf{k}}v_{\mathbf{k}}v_{\mathbf{q}}^{2}(\hat{b}_{\mathbf{k}}%
\hat{b}_{-\mathbf{k}}+\hat{b}_{\mathbf{k}}^{+}\hat{b}_{-\mathbf{k}}^{+})\hat{%
b}_{\mathbf{q}}\hat{b}_{\mathbf{q}}^{+}+u_{\mathbf{k}}v_{\mathbf{k}}u_{%
\mathbf{q}}v_{\mathbf{q}}(\hat{b}_{\mathbf{k}}\hat{b}_{-\mathbf{k}}+\hat{b}_{%
\mathbf{k}}^{+}\hat{b}_{-\mathbf{k}}^{+})(\hat{b}_{\mathbf{q}}\hat{b}_{-%
\mathbf{q}}+\hat{b}_{\mathbf{q}}^{+}\hat{b}_{-\mathbf{q}}^{+})\right] .
\end{equation}%
To calculate the expectation value of the terms with four quasiparticle
operators appearing in $<\hat{N}_{out}^{2}>$ we use Wick's theorem which
holds for the operators of statistically independent excitations. In
particular, nonzero averages come from the terms with $\mathbf{q}=\pm 
\mathbf{k}$. Using Wick's theorem we obtain%
\[
<\hat{b}_{\mathbf{k}}^{+}\hat{b}_{\mathbf{k}}\hat{b}_{\mathbf{k}}^{+}\hat{b}%
_{\mathbf{k}}>=2f_{k}^{2}+f_{k},\qquad <\hat{b}_{\mathbf{k}}^{+}\hat{b}_{%
\mathbf{k}}\hat{b}_{-\mathbf{k}}^{+}\hat{b}_{-\mathbf{k}}>=f_{k}^{2}, 
\]%
\[
<\hat{b}_{\mathbf{k}}\hat{b}_{\mathbf{k}}^{+}\hat{b}_{\mathbf{k}}\hat{b}_{%
\mathbf{k}}^{+}>=2f_{k}^{2}+3f_{k}+1,\qquad <\hat{b}_{\mathbf{k}}\hat{b}_{%
\mathbf{k}}^{+}\hat{b}_{-\mathbf{k}}\hat{b}_{-\mathbf{k}%
}^{+}>=(f_{k}+1)^{2}, 
\]%
\[
<\hat{b}_{\mathbf{k}}^{+}\hat{b}_{\mathbf{k}}\hat{b}_{\mathbf{k}}\hat{b}_{%
\mathbf{k}}^{+}>=<\hat{b}_{\mathbf{k}}\hat{b}_{\mathbf{k}}^{+}\hat{b}_{%
\mathbf{k}}^{+}\hat{b}_{\mathbf{k}}>=2f_{k}^{2}+2f_{k}, 
\]%
\[
<\hat{b}_{\mathbf{k}}^{+}\hat{b}_{\mathbf{k}}\hat{b}_{-\mathbf{k}}\hat{b}_{-%
\mathbf{k}}^{+}>=<\hat{b}_{-\mathbf{k}}\hat{b}_{-\mathbf{k}}^{+}\hat{b}_{%
\mathbf{k}}^{+}\hat{b}_{\mathbf{k}}>=f_{k}^{2}+f_{k}, 
\]%
\[
<\hat{b}_{\mathbf{k}}\hat{b}_{-\mathbf{k}}\hat{b}_{\mathbf{k}}^{+}\hat{b}_{-%
\mathbf{k}}^{+}>=<\hat{b}_{\mathbf{k}}\hat{b}_{-\mathbf{k}}\hat{b}_{\mathbf{%
-k}}^{+}\hat{b}_{\mathbf{k}}^{+}>=(f_{k}+1)^{2}, 
\]%
\[
<\hat{b}_{\mathbf{k}}^{+}\hat{b}_{-\mathbf{k}}^{+}\hat{b}_{\mathbf{k}}\hat{b}%
_{-\mathbf{k}}>=<\hat{b}_{\mathbf{k}}^{+}\hat{b}_{-\mathbf{k}}^{+}\hat{b}_{%
\mathbf{-k}}\hat{b}_{\mathbf{k}}>=f_{k}^{2}, 
\]
and therefore%
\[
<\hat{N}_{out}^{2}>=\sum\limits_{\mathbf{k}\neq \pm \mathbf{q}\neq 0}\left[
u_{\mathbf{k}}^{2}u_{\mathbf{q}}^{2}f_{k}f_{q}+v_{\mathbf{k}}^{2}v_{\mathbf{q%
}}^{2}(f_{k}+1)(f_{q}+1)+2u_{\mathbf{k}}^{2}v_{\mathbf{q}}^{2}f_{k}(f_{q}+1)%
\right] + 
\]%
\[
\sum\limits_{k\neq 0}\left[ u_{\mathbf{k}}^{4}(<\hat{b}_{\mathbf{k}}^{+}\hat{%
b}_{\mathbf{k}}\hat{b}_{\mathbf{k}}^{+}\hat{b}_{\mathbf{k}}>+<\hat{b}_{%
\mathbf{k}}^{+}\hat{b}_{\mathbf{k}}\hat{b}_{-\mathbf{k}}^{+}\hat{b}_{\mathbf{%
-k}}>)+v_{\mathbf{k}}^{4}(<\hat{b}_{\mathbf{k}}\hat{b}_{\mathbf{k}}^{+}\hat{b%
}_{\mathbf{k}}\hat{b}_{\mathbf{k}}^{+}>+<\hat{b}_{\mathbf{k}}\hat{b}_{%
\mathbf{k}}^{+}\hat{b}_{\mathbf{-k}}\hat{b}_{\mathbf{-k}}^{+}>)\right. + 
\]%
\[
u_{\mathbf{k}}^{2}v_{\mathbf{k}}^{2}\left( <\hat{b}_{\mathbf{k}}^{+}\hat{b}_{%
\mathbf{k}}\hat{b}_{\mathbf{k}}\hat{b}_{\mathbf{k}}^{+}>+<\hat{b}_{\mathbf{k}%
}\hat{b}_{\mathbf{k}}^{+}\hat{b}_{\mathbf{k}}^{+}\hat{b}_{\mathbf{k}}>+<\hat{%
b}_{\mathbf{k}}\hat{b}_{-\mathbf{k}}\hat{b}_{\mathbf{k}}^{+}\hat{b}_{-%
\mathbf{k}}^{+}>+<\hat{b}_{\mathbf{k}}^{+}\hat{b}_{-\mathbf{k}}^{+}\hat{b}_{%
\mathbf{k}}\hat{b}_{-\mathbf{k}}>\right. + 
\]%
\[
\left. \left. <\hat{b}_{\mathbf{k}}^{+}\hat{b}_{\mathbf{k}}\hat{b}_{\mathbf{%
-k}}\hat{b}_{\mathbf{-k}}^{+}>+<\hat{b}_{\mathbf{-k}}\hat{b}_{\mathbf{-k}%
}^{+}\hat{b}_{\mathbf{k}}^{+}\hat{b}_{\mathbf{k}}>+<\hat{b}_{\mathbf{k}}\hat{%
b}_{-\mathbf{k}}\hat{b}_{\mathbf{-k}}^{+}\hat{b}_{\mathbf{k}}^{+}>+<\hat{b}_{%
\mathbf{k}}^{+}\hat{b}_{-\mathbf{k}}^{+}\hat{b}_{\mathbf{-k}}\hat{b}_{%
\mathbf{k}}>\right) \right] = 
\]%
\[
\sum\limits_{\mathbf{k}\neq \pm \mathbf{q}\neq 0}\left[ u_{\mathbf{k}}^{2}u_{%
\mathbf{q}}^{2}f_{k}f_{q}+v_{\mathbf{k}}^{2}v_{\mathbf{q}%
}^{2}(f_{k}+1)(f_{q}+1)+2u_{\mathbf{k}}^{2}v_{\mathbf{q}}^{2}f_{k}(f_{q}+1)%
\right] + 
\]%
\begin{equation}
\sum\limits_{k\neq 0}\left[ u_{\mathbf{k}}^{4}(3f_{k}^{2}+f_{k})+v_{\mathbf{k%
}}^{4}(3f_{k}^{2}+5f_{k}+2)+u_{\mathbf{k}}^{2}v_{\mathbf{k}%
}^{2}(10f_{k}^{2}+10f_{k}+2)\right] .  \label{b3}
\end{equation}%
From the other hand 
\begin{equation}
<\hat{N}_{out}>^{2}=\sum\limits_{k\neq 0}\sum\limits_{q\neq 0}\left[ \left(
u_{k}^{2}+v_{k}^{2}\right) f_{k}+v_{k}^{2}\right] \left[ \left(
u_{q}^{2}+v_{q}^{2}\right) f_{q}+v_{q}^{2}\right] .  \label{b4}
\end{equation}%
Using the particle number constraint together with Eqs. (\ref{b3}) and (\ref%
{b4}) and $u_{\mathbf{k}}^{2}-v_{\mathbf{k}}^{2}$ $=1$ we find the following
answer for the squared variance of condensate fluctuations [Giorgini et al
(1998)]%
\[
\mu _{2}\equiv <\hat{n}_{0}^{2}>-<\hat{n}_{0}>^{2}=<\hat{N}_{out}^{2}>-<\hat{%
N}_{out}>^{2}= 
\]%
\begin{equation}
\sum\limits_{k\neq 0}\left[
(1+8u_{k}^{2}v_{k}^{2})(f_{k}^{2}+f_{k})+2u_{k}^{2}v_{k}^{2}\right] .
\label{b5}
\end{equation}

Abrikosov, A.A., Gorkov, L.P. and Dzyaloshinskii, I.E. (1963) \textit{%
Methods of Quantum Field Theory in Statistical Physics} (Prentice-Hall,
Englewood Cliffs, N.J.).

Anderson, M. Ensher, J., Matthews, M., Wieman, C. and Cornell, E. (1995), 
\textit{Science} 269, 198.

Anderson, B.P. and Kasevich, M.A. (1998), \textit{Science} 282, 1686.

Anderson, B.P. and Kasevich, M.A. (1999), \textit{Phys. Rev. A} \textbf{\ }%
59, R938.

Andrews, M.R., Townsend, C.G., Miesner, H.J., Durfee, D.S., Kurn, D.M. and
Ketterle, W. (1997), \textit{Science} 275, 637.

Baker, G.A. and Graves-Morris, P. (1996), Pad\'{e} Approximants. New York:
Cambridge University Press.

Bloch, I., H\"{a}nsch, T.W. and Esslinger, T. (1999), \textit{Phys. Rev.
Lett.} 82, 3008.

Bohr, N. (1913), ``On the constitution of atoms and molecules (Part 1)", 
\textit{Philosophical Magazine}, 26, 1.

Boltzmann, L. (1877a), ``\"{U}ber die Beziehung zwischen dem zweiten
Hauptzatz der mechanischen W\"{a}rmetheorie und der
Warhscheinlichkeitsrechnung, respective den S\"{a}tzen \"{u}ber dan W\"{a}%
rmegleichgewicht in Wissenshaftliche Abhandlungen", \textit{Wiener Bericte},
2 (76): \ 373-435. Reprinted in Brush, 2007.

Boltzmann, L. (1877b), `` \"{U}ber die Beziehung eines allgemeine
mechanischen Satzes zum zweiten Hauptsatze der W\"{a}rmetheorie",
Sitzungsberichte Akad. Wiss., Bienna, part II, 75, 67-73. Also reprinted in
Brush (2003). This is Boltzmann's restatement of and refutation of
Loschmidt's reversibility argument.

Boltzmann, L. (1884), ``Ableitung des Stefan'schen Gesetzes etereffend die
abh\"{a}ngigkeit der W\"{a}rmestrahlung von der Temperature aus der
electromagnetischen Lichttheorie", \textit{Annalen der Physik} 22, 291.

Boltzmann, L. (1896a), ``Entgegnung auf die wtheoretischen Betrachtungen des
Hrn. E. Zermelo", Annalen der Physik 57, 773-784. This and Boltzmann (1896b)
are Boltzmann's answers to Zermelo (1696a) and Zermelo (1896b). All four
papers are reprinted in Brush (2007).

Boltzmann, L. (1896b), ``Zu Hrn. Zermelo's Abhandlung \"{U}ber die
mechanische Erkl\"{a}rung irreversibler Vorgange", Annalen der Physik 60,
392-398.

Born (1949)], \textit{Atomic Physics}. Fifth Edition, Hafner, New York
(1949). Reprinted by Dover, New York, (1989).

Bradley, C., Sackett, C., Tollett, J. and Hulet, R. (1995), \textit{Phys.
Rev. Lett.} 75, 1687.

Brush, S. G.(2003), Hall, N. S. (2003), ed., ``\textit{The Kinetic Theory of
Gases: An Anthology of Classic Papers with Historical Commentary}", Imperial

Chan, M. H. W., Blum, K.I., Murphy, S.Q., Wong, G. K. S. and Reppy, J.D.
(1988), \textit{Phys. Rev. Lett.} 61, 1950.

Chuu, C.S., Schreck, F., Meyrath, T.P., Hanssen, J.L., Price, G.N. and
Raizen, M.G. (2005), \textit{Phys. Rev. Lett.} 95, 260403. This paper
reports the recent experiment of M. Raizen and coworkers on observation of
number statistics in a degenerate Bose gas.

College Press, London. This contains reprints and commentaries on the papers
Boltzmann (1877a,b), Boltzmann (1896a,b), Zermelo (1896a,b).

Crooker, B.C., Hebral, B., Smith, E.N., Takano, Y. and Reppy, J.D. (1983), 
\textit{Phys. Rev. Lett.} 51, 666.

Crowell, P.A., Van Keuls, F.W. and Reppy, J.D. (1995), \textit{Phys. Rev.
Lett.} 75, 1106.

Davis, K., Mewes, M., Andrews, M., van Druten, N., Durfee, D., Kurn, D. and
Ketterle, W. (1995), \textit{Phys. Rev. Lett.} 75, 3969.

DeGiorgio, V. and Scully, M.O. (1970), \textit{Phys. Rev.} A2, 1170.

Dulong, P. L. and Petit, A. T. (1819), ``Sur quelques points important s de
la th\'{e}erie de la chaleur", \textit{Annales de Chimie et de Physique }10,
395-413.

Ehrenfest, P. (1906), ``Planckschen Strahlungstheorie", Physikalische
Zeitschrift 7, 528-532. Reprinted in his collected papers, p. 123, (see
Klein (1959)).

Ehrenfest, P. (1911), ``Welsche Z\"{u}ge der Lichtquantenhypothesese spielen
in der Theorie der W\"{a}rmestrahlung eine wesentlische Role." \textit{%
Annalen der Physik}, 36:91-118.

Ehrenfest, P. (1913), This article was published in Dutch. It is reprinted
in english translation in as ``A mechanical theorem of Boltzmann and its
relation to the theory of energy quanta", Proceedings of the Amsterdam
Academy 16, 591-597 (1914); and reprinted in his collected papers, p.
340-346, (see Klein (1959)).

Einstein, A (1905), `` \"{U}ber einen die Erzeugung und Verwandlung des
Lichtes betreffenden heuristischen Gesichspunkt". \textit{Annalen der Physik}%
, 17:132-148.

Einstein, A. (1907), ``Die Plancksche Theorie der Strahlung und die Theorie
der spezifischen W\"{a}rme". \textit{Annalen der Physik}, 22:180-190.

Einstein, A. (1909), ``On the present status of the radiation problem." 
\textit{Physikalische Zeitschrift}, 10:185-193.

Einstein, A. (1917), ``Zur Quantentheorie der Strahlung", \textit{%
Physicalische Zeitschrift}, 18:121-128. Reprinted in english translation in
Ter Haar (1967).

Einstein, A. (1924), \emph{Sitzungsberichte der preussischen Akademie der
Wissenschaften\/}, XXII.~Gesamtsitzung, p.~261.

Einstein, A. (1925), \emph{Sitzungsberichte der preussischen Akademie der
Wissenschaften\/}, I.~Sitzung der physikalisch-mathematischen Klasse, p.~3.

Ernst, U., Marte, A., Schreck, F., Schuster, J. and Rempe, G. (1998), 
\textit{Europhys. Lett. }41, 1.

Esslinger, T., Bloch, I. and Hansch, T.W. (1998), \textit{Phys. Rev. A} 58,
R2664.

Fetter, A.L. and Walecka, J.D. (1971), \textit{Quantum Theory of
Many-Particle Systems} (McGraw-Hill, San Francisco).

Fierz, M. (1956), \textit{Helv. Phys. Acta} 29, 47.

Fried, D.G., Killian, T.C., Willmann, L., Landhuis, D., Moss, S.C.,
Kleppner, D. and Greytak, T.J. (1998), \textit{Phys. Rev. Lett.} 81, 3811.

Gajda, M. and Rz\c{a}\.{z}ewski, K. (1997), \textit{Phys. Rev. Lett.} 78,
2686.

Giorgini, S., Pitaevskii, L.P. and Stringari, S. (1998), \textit{Phys. Rev.
Lett.} 80, 5040

Glauber, R. J. (1964), \textquotedblleft Quantum Optics and Electronics," p.
162, Les Houches Summer Lectures, DeWitt, C., Blandin, A. and
Cohen-Tannoudji, C. eds.

Graham, R. and Haken, H. (1970), \textit{Z. Physik} 237, 31.

Grossmann, S. and Holthaus, M. (1996), \textit{Phys. Rev. E} 54, 3495.

Han, D.J., Wynar, R.J., Courteille, Ph. and Heinzen, D.J. (1998), \textit{%
Phys. Rev. A} 57, R4114.

Hau, L.V., Busch, B.D., Liu, C., Dutton, Z., Burns, M.M. and Golovchenko,
J.A. (1998), \textit{Phys. Rev. A} 58, R54.

Heilbron, J. L. (1996), ``\textit{The dilemmas of an upright man: Max Planck
and the fortunes of German science}", Harvard Univ. Press, Cambridge.

Hermann, A. (1971), ``\textit{The genesis of quantum theory (1899-1913)}".
MIT Press, Cambridge, Mass.

Idziaszek, Z., Rz\c{a}\.{z}ewski, K. and Lewenstein, M. (2000), \textit{%
Phys. Rev. A} 61, 053608.

Jammer, M. (1966), ``\textit{The Conceptual Development of Quantum Mechanics}%
", McGraw-Hill, New York.

Jeans, J. H. (1905), ``On the partition of energy between matter and ether", 
\textit{Philosophical Magazine} 10, 91-98.

Jordan, A. N., Ooi, C. H. R. and Svidzinsky, A.A. (2006), \textit{Phys. Rev.
A} 74, 032506.

Kangro, H. (1976), ``\textit{Early History of Planck's Radiation Law}",
Taylor \& Francis, London.

Kapale, K.T. and Zubairy, M.S. (2001), \textit{Opt. Commun.} 191, 299.

Ketterle, W. and van Druten, N. J. (1996), \textit{Phys. Rev. A}\textbf{\ }%
54, 656.

Kirchoff, G. (1860), ``\"{U}ber den Zusammenhang von Emission und Absorption
von Licht und W\"{a}rme", \textit{Monatsberichte der Akademie der} \textit{%
Wissenschaften zu Berlin}, 1859:783-787.

Klein, M. J., ed., (1959), \textit{Paul Ehrenfest, Collected Scientific
Papers}, North-Holland Publishing Co., Amsterdam, 340-346, (1959).

Klein, M. (1975), ``Max Planck and the beginnings of the quantum theory". 
\textit{Archive for History of Exact Sciences}, 1(5):459-479.

Kocharovsky, V.V., Scully, M.O., Zhu, S.Y. and Zubairy, M.S. (2000a), 
\textit{Phys. Rev. A} 61, 023609.

Kocharovsky, V.V., Kocharovsky, Vl.V. and Scully, M.O. (2000b), \textit{%
Phys. Rev. Lett. }84, 2306; \textit{Phys. Rev. A }61\textit{,} 053606.

Kocharovsky, V.V., Kocharovsky, Vl.V., Holthaus, M., Ooi, C.H.R.,
Svidzinsky, A.A., Ketterle, W. and Scully, M.O. (2006), \textit{Ad. in AMO
Physics}, 53, 291.

Kuhn, T. S. (1978), ``\textit{Blackbody Radiation and the Quantum
Discontinuity 1894-1912}". Reprinted (1987), University of Chicago, Chicago,
with an added Appendix, ``\textit{Planck Revisited}". Kuhn takes exception,
as we do, to the notion that Planck quantized anything physical.

Leggett, A. (2001), \textit{Rev. Mod. Phys.} 73, 307.

Lewis, G. N. (1926), ``The conservation of photons". Nature 118, 874-875. In
this paper Lewis introduced the name ``photon".

Lifshitz, E.M. and Pitaevskii, L.P. (1981), \textit{Statistical Physics,
Part 2} (Pergamon, Oxford).

Lorentz, H. A. (1908), His famous report to the Rome Academy, supporting the
Rayleigh-Jeans Law. Reprinted in improved form in Nuevo Cimento, 16, 5-34,
(1908).

Lorentz, H. A. (1912), ``Les th\'{e}ories statistiques en thermodynamique", 
\textit{Conf\'{e}rences faites au Coll\`{e}ge de France en Novembre 1912},
Teubner, Leipzig, (1916).

Mehra, J. and Rechenberg, H. (1982), ``\textit{The Historical Development of
Quantum Theory"}, Vol.1, Part 1, Springer-Verlag, New York.

Mewes, M.O., Andrews, M.R., Kurn, D.M., Durfee, D.S., Townsend, C.G. and
Ketterle, W. (1997), \textit{Phys Rev. Lett.} 78, 582;

Miesner, H.J., Stamper-Kurn, D., Andrews, M., Durfee, D., Inouye, S. and
Ketterle, W. (1998), \textit{Science} 279, 1005.

Natanson, L. (1911), ``\"{U}ber die statistiche Theorie der Strahlung", 
\textit{Physicalische Zeitschrift}, 12: 659-666.

Planck, M. (1900a), ``\"{U}ber irreversible Strahlungsvorg\"{a}nge", \textit{%
Annalen der Physik}, 1:69. Reprinted in english translation in Ter Haar
(1967).

Planck, M. (1900b). ``Zur Theorie des Gesetzes der Energieverteilung im
Normalspektrum", \textit{Verhandlungen der Deutschen Physikalischen
Gesellschaft}, 2(17): 237-245. Reprinted in english translation in Ter Haar
(1967).

Planck M. (1900c), \"{U}ber irreversible Strahlungsvorg\"{a}nge", \textit{%
Annalen der Physik }1, 69-122. These are Planck's thoughts on rederiving
Wien's spectral law.

Planck, M. (1900d), \textit{Annalen der Physik}, 1, 730.

Planck, M. (1913), ``\textit{Theory of Heat Radiation}", second edition,
english translation put out by Dover Publ., New York, (1959). Reprinted
(1991).

Planck (1931). These comments appear in a famous upublished letter to R. W.
Wood. The letter is on file at the Center for the History and

Politzer, H. D. (1996), \textit{Phys. Rev. A} 54, 5048.

Philosophy of Physics, at the American Institute of Physics, College Park,
Md. \ The letter is quoted in full in Hermann (1971).

Lord Rayleigh, (1900), \textit{Pilosophical Magazine} 49, 530-540.

Lord Rayleigh, (1905), ``The Dynamical Theory of Gases and Radiation", 
\textit{Nature} 72, 54-55.

Rosenfeld, L. (1936), ``La premi\`{e}re phase de l'evolution de la th\'{e}%
orie des quanta", \textit{Osiris} 2, 149-196.

Scully, M.O. and Lamb, W.E. (1966), \textit{Phys. Rev. Lett.} 16, 853.

Scully, M.O. and Zubairy, M.S. (1997), \textit{Quantum Optics} (Cambridge,
London).

Scully, M.O. (1999), \textit{Phys. Rev. Lett.} 82, 3927.

Shi, H. and Griffin, A. (1998), \textit{Phys. Rep.} 304, 1.

Stefan, J. (1879), ``\"{U}ber die Beziehung zwischen der W\"{a}rmestrahlung
und der Temperatur", \textit{Wien. Akad. Sitzber}., 79:391-428.

Svidzinsky, A.A. and Scully, M.O. (2006), \textit{Phys. Rev. Lett. }97,
190402.

Ter Haar, D. (1967), ``\textit{The Old Quantum Theory}", Pergamon Press,
Oxford. This reprint collection has notes on and reprints of Planck
(1900a,b), and Einstein (1917).

Ter Haar, D. (1970), \emph{Lectures on selected topics in statistical
mechanics} (Elsevier, Amsterdam), Chapter 4; Fujiwara, I, ter Haar, D. and
Wergeland, H. (1970), \textit{J. Stat. Phys. }2, 329.

Varro, S. (2006), Einstein's Fluctuation Formula, a Historical Overview, 
\textit{Fluctuation and Noise Letters}, Vol. 6, No. 3, R11-R46. A critical
review of fluctuations in Einstein's and Bose's work.

Wien, W. (1893), Sitzungsber. der Berlin Akad., 1893. \textit{Annalen der
Physick (Leipzig)} 52: 132-165.

Wien, W. (1896), ``\"{U}ber die Energievertheilung im Emissionsspectrum
eines schwarzen K\"{o}rpers", \textit{Wiedemannsche Annalen der Physik} 58,
662-669.

Wilkens, M. and Weiss, C. (1997), \textit{J. Mod. Opt.} 44, 1801.

Weiss, C. and Wilkens, M. (1997), \textit{Optics Express} 1, 272.

Zermelo, E. (1896a), ``\"{U}ber enen Satz der Dynamik und die mechanische W%
\"{a}rmetheorie", \textit{Annalen der Physik} 57, 485-494. Answered by
Boltzmann in Boltzmann (1896). Zermelo's two papers and Boltzmann's answers
are both reprinted in Brush (2003).

Zermelo, E. (1896b)', ``\"{U}ber mechanische Erkl\"{a}rungen irreversibler
Vorg\"{a}nge", \textit{Annalen der Physik} 59, 793-801.

Ziff, R.M., Uhlenbeck, G.E. and Kac, M. (1977), \textit{Phys. Rep. }32, 169.

\end{document}